\documentclass[a4paper,aps,prd,10pt,preprintnumbers,showpacs,twocolumn,superscriptaddress,nofootinbib,amsmath,amssymb]{revtex4-2}
\usepackage[utf8]{inputenc}
\usepackage[T1]{fontenc}
\usepackage{cmap}
\usepackage{graphicx}
\usepackage{tabularx}
\usepackage{makecell} 
\usepackage{makecell}
\usepackage[utf8]{inputenc}
\usepackage[T1]{fontenc}
\usepackage{amsfonts}
\usepackage{amsmath}
\usepackage{amssymb}
\usepackage{amsthm}
\usepackage{mathrsfs}
\usepackage{euscript}
\usepackage{color}
\usepackage{braket}
\usepackage{tensor}
\usepackage{mathtools}
\usepackage{float} 
\usepackage{longtable}
\usepackage[colorlinks=true,
linkcolor=blue,
citecolor=blue,
urlcolor=blue]{hyperref}

\begin{document}
	
	\title{Radial spectra and dynamical signatures of excited boson stars}
	
	\author{Chen-Hao Hao}
	\affiliation{Department of Physics, Key Laboratory of Low Dimensional Quantum Structures and Quantum Control of Ministry of Education, and Synergetic Innovation Center for Quantum Effects and Applications, Hunan Normal University, Changsha, Hunan 410081, P. R. China}
	\affiliation{Institute of Interdisciplinary Studies, Hunan Normal University, Changsha, Hunan 410081, P. R. China}

	\author{Wen-Di Guo}
	\affiliation{Lanzhou Center for Theoretical Physics, Key Laboratory of Theoretical Physics of Gansu Province, Key Laboratory of Quantum Theory and Applications of MoE, Gansu Provincial Research Center for Basic Disciplines of Quantum Physics, Lanzhou University, Lanzhou 730000, China}
	\affiliation{Institute of Theoretical Physics \& Research Center of Gravitation, School of Physical Science and Technology, Lanzhou University, Lanzhou 730000, China}
	
	\author{Qin Tan}
	\email{tanqin@hunnu.edu.cn, corresponding author}
	\affiliation{Department of Physics, Key Laboratory of Low Dimensional Quantum Structures and Quantum Control of Ministry of Education, and Synergetic Innovation Center for Quantum Effects and Applications, Hunan Normal University, Changsha, Hunan 410081, P. R. China}
	\affiliation{Institute of Interdisciplinary Studies, Hunan Normal University, Changsha, Hunan 410081, P. R. China}
	
	\author{Jieci Wang}
	\email{jcwang@hunnu.edu.cn, corresponding author}
	\affiliation{Department of Physics, Key Laboratory of Low Dimensional Quantum Structures and Quantum Control of Ministry of Education, and Synergetic Innovation Center for Quantum Effects and Applications, Hunan Normal University, Changsha, Hunan 410081, P. R. China}
	\affiliation{Institute of Interdisciplinary Studies, Hunan Normal University, Changsha, Hunan 410081, P. R. China}
	
	\begin{abstract}
		We compute the lowest radial mode of spherically symmetric boson stars along equilibrium branches with a fixed number of radial nodes, considering both mini boson stars and quartically self-interacting models. By reformulating the pulsation equations in additive variables that remain regular at the zeros of the background scalar field, the eigenvalue problem can be integrated directly through the nodes of excited configurations. For all branches examined, the first zero of the constrained fundamental radial eigenvalue coincides, within numerical resolution, with the first simultaneous critical point of the Arnowitt--Deser--Misner (ADM) mass, Noether charge, and binding energy. We further evaluate the radial eigenvalue for the threshold models identified in nonlinear spherical evolutions of excited boson stars and find a simple empirical correlation with the node number and self-interaction strength. Our results provide a regular perturbative framework for  excited boson stars and clarify the relation between constrained radial modes, equilibrium critical points, and nonlinear stability diagnostics.
	\end{abstract}

	\maketitle
	
	\section{Introduction}
	\label{sec:introduction}
	
	The detection of gravitational waves and the horizon-scale images reported by the Event Horizon Telescope have greatly strengthened the astrophysical case for black holes \cite{LIGOScientific:2016aoc,LIGOScientific:2018mvr,
		EventHorizonTelescope:2019dse,EventHorizonTelescope:2022wkp}.  Nevertheless, compact horizonless objects remain important as controlled alternatives for testing strong field gravity and, in suitable regimes, can mimic some observational signatures of black holes.  Among the simplest and best studied examples are boson stars: self-gravitating configurations of bosonic fields, most commonly modeled as classical solutions of the Einstein-Klein-Gordon system for a complex scalar field minimally coupled to gravity.  Depending on the scalar mass and self-interaction potential, boson stars have been regarded as exotic compact objects, dark matter configurations, and bosonic components of mixed systems \cite{Kaup:1968zz,Ruffini:1969qy,Schunck:2003kk,Liebling:2012fv,
		Vincent:2015xta,Olivares:2018abq,Palenzuela:2017kcg,Cardoso:2019rvt,
		Lee:1995af,Arbey:2003sj,Liang:2022mjo,Ma:2023vfa,Hao:2025gak,
		Huang:2026bgc,Sun:2022duv,Hao:2024hba,Herdeiro:2021lwl,Herdeiro:2026agu}. In the simplest spherical setting, the geometry is static whereas the scalar field has harmonic time dependence, $\phi(t,r)=\phi_0(r)e^{-i\omega t}$, with $\omega$ acting as an eigenvalue of the self-gravitating configuration.		
	An equilibrium solution alone does not establish physical viability; stability under perturbations and evolution is essential. For boson stars, complementary approaches include linear
	radial perturbations, nonlinear time evolutions, and critical-point
	arguments. Radial perturbation theory identifies oscillatory and unstable modes through the sign of the squared eigenfrequency
	\cite{Gleiser:1988rq,Jetzer:1988vr,Gleiser:1988ih,Lee:1988av,
		Alcubierre:2021mvs,Kojima:1991np,Yoshida:1994xi,Macedo:2013jja,
		Visinelli:2021uve,Cardoso:2020nst,Tan:2022uex,Tan:2024hzw}, whereas evolutions determine possible outcomes such as migration, dispersal, and collapse
	\cite{Seidel:1990jh,Balakrishna:1997ej,Hawley:2000dt,Nyhan:2022pda,
		DiGiovanni:2022xcc,Zhang:2022qzw,Zhang:2023qxf,Ge:2024itl}.
	Critical-point methods instead diagnose candidate stability changes through extrema of global quantities such as the ADM mass, Noether charge, and binding energy \cite{Kusmartsev:1990cr,Kleihaus:2011sx}. For the standard
	nodeless branch, the first mass maximum coincides with the zero of the fundamental radial mode \cite{Kain:2021rmk}, although such a correspondence does not hold universally \cite{Santos:2024vdm}.
	
	On the other hand, excited states provide a general probe of the internal organization and dynamical response of bound systems, since their spectra and decay channels reveal how microscopic structure controls macroscopic stability. For
	self-gravitating configurations, this issue is especially nontrivial because matter and spacetime geometry form a nonlinear, self-consistent eigenstate.
	Excited boson stars therefore offer a controlled setting in which to
	investigate how radial quantum structure affects compact object dynamics. Their equilibrium scalar profiles possess radial nodes: the ground state has no node, while the $n-$th excited background has $n$ nodes. A nodeful configuration may be viewed as a macroscopic occupation of a radially excited
	bosonic level rather than the lowest level. Dynamical studies show that excited mini boson stars are generically unstable and may migrate to lower-node configurations or collapse, whereas sufficiently strong quartic self-interactions can render excited spherical configurations long lived under spherical evolution
	\cite{Sanchis-Gual:2021phr,Brito:2023fwr}. These results raise a broader question: how are node structure, equilibrium critical points, and dynamical stability encoded in the linear response of the system?
	
	Radial perturbations provide a direct spectral route to this question.
	Earlier studies established stability criteria for excited boson stars under radial perturbations \cite{Jetzer:1989vs}, but the fundamental mode across nodeful branches and its relation to equilibrium and dynamical diagnostics remain insufficiently understood. In this paper, we develop a regular radial
	perturbation formulation applicable across the nodes of excited backgrounds and determine their fundamental radial modes. Following each branch as a function of the scalar frequency $\omega$, we compare the zero crossing of $\chi_{0,n}^2$ with the critical points of the ADM mass, Noether charge, and binding energy. By further connecting the spectrum with nonlinear stability
	thresholds, we test whether the linear response around a static equilibrium can retain information about its time-dependent dynamics.
	
	The remainder of the paper is organized as follows.  Section~\ref{sec:equilibrium} describes the equilibrium branches and turning-point diagnostics.  Section~\ref{sec:perturbation} presents the regular radial perturbation system, shooting method, and zero-mode diagnostic.  Results and conclusions are presented in Secs.~\ref{sec:results} and \ref{sec:conclusions}, respectively.
	
	\section{Equilibrium configurations}
	\label{sec:equilibrium}
	
	We label an equilibrium branch by the number $n$ of radial nodes of the real scalar profile $\phi_0(r)$.  Thus $n=0$ denotes the ground state branch and $n\geq1$ denotes an excited branch.  The fundamental radial eigenvalue on the $n$th background branch is denoted by $\chi_{0,n}^2$.  Throughout this work we use units $c=\hbar=1$ and metric signature $(-+++)$. 
	
	\subsection{Einstein-Klein-Gordon system}
	\label{subsec:ekg_system}
	
	We consider a complex scalar field minimally coupled to gravity \cite{Kaup:1968zz,Ruffini:1969qy},
	\begin{equation}
		S=\int d^4x\sqrt{-g}
		\left[
		\frac{R}{16\pi G}
		-g^{\mu\nu}\partial_\mu\phi^\ast\partial_\nu\phi
		-V(|\phi|^2)
		\right],
		\label{eq:action}
	\end{equation}
	with quartic potential
	\begin{equation}
		V(|\phi|^2)=\mu^2|\phi|^2+\eta|\phi|^4.
		\label{eq:potential}
	\end{equation}
	Defining $s\equiv|\phi|^2$, one can write
	\begin{equation}
		W(s)\equiv\frac{dV}{ds}=\mu^2+2\eta s,
		\qquad
		W_s(s)\equiv\frac{dW}{ds}=2\eta.
		\label{eq:W_def}
	\end{equation}
	The spacetime is assumed to be spherically symmetric,
	\begin{equation}
		ds^2=-e^{\nu(t,r)}dt^2+e^{\lambda(t,r)}dr^2+r^2d\Omega^2.
		\label{eq:metric}
	\end{equation}
	The Einstein equations are
	\begin{equation}
		G^\mu{}_{\nu}=8\pi G T^\mu{}_{\nu},
		\label{eq:einstein}
	\end{equation}
	with
	\begin{equation}
		T_{\mu\nu}=
		\partial_\mu\phi^\ast\partial_\nu\phi
		+\partial_\nu\phi^\ast\partial_\mu\phi
		-g_{\mu\nu}
		\left(g^{\alpha\beta}\partial_\alpha\phi^\ast\partial_\beta\phi+V\right).
		\label{eq:Tmunu}
	\end{equation}
	For the metric~\eqref{eq:metric}, the $(t,t)$ and $(r,r)$ equations may be written as
	\begin{align}
		\nu'&=8\pi G r e^\lambda T^r{}_r+\frac{e^\lambda-1}{r},
		\label{eq:nu_general}\\
		\lambda'&=-8\pi G r e^\lambda T^t{}_t-\frac{e^\lambda-1}{r}.
		\label{eq:lambda_general}
	\end{align}
	A prime and an overdot denote derivatives with respect to $r$ and $t$, respectively.  The scalar field equation is
	\begin{equation}
		\nabla_\mu\nabla^\mu\phi-W(|\phi|^2)\phi=0.
		\label{eq:KG_general}
	\end{equation}
	The global $U(1)$ symmetry gives the conserved current
	\begin{equation}
		J^\mu=i g^{\mu\nu}
		\left(\phi\,\partial_\nu\phi^\ast-\phi^\ast\partial_\nu\phi\right).
		\label{eq:current}
	\end{equation}

	\subsection{Ground-state and excited branches}
	\label{subsec:equilibrium_solutions}
	
	A static geometry is obtained with \cite{Schunck:2003kk}
	\begin{equation}
		\phi(t,r)=\phi_0(r)e^{-i\omega t},
		\label{eq:static_ansatz}
	\end{equation}
	where $\omega$ is real and $\phi_0$ may be chosen real.  The static Klein-Gordon equation is
	\begin{equation}
		\phi_0''+
		\left(\frac{2}{r}+\frac{\nu_0'-\lambda_0'}{2}\right)\phi_0'
		+e^{\lambda_0}
		\left[\omega^2e^{-\nu_0}-W(\phi_0^2)\right]\phi_0=0.
		\label{eq:static_KG}
	\end{equation}
	The relevant stress-energy components are
	\begin{align}
		T^t{}_t&=-\omega^2e^{-\nu_0}\phi_0^2-e^{-\lambda_0}{\phi_0'}^2-V(\phi_0^2),
		\label{eq:Ttt_static}\\
		T^r{}_r&=+\omega^2e^{-\nu_0}\phi_0^2+e^{-\lambda_0}{\phi_0'}^2-V(\phi_0^2).
		\label{eq:Trr_static}
	\end{align}
	The metric equations are
	\begin{align}
		\nu_0'&=8\pi G r e^{\lambda_0}T^r{}_r+\frac{e^{\lambda_0}-1}{r},
		\label{eq:nu_static}\\
		\lambda_0'&=-8\pi G r e^{\lambda_0}T^t{}_t-\frac{e^{\lambda_0}-1}{r}.
		\label{eq:lambda_static}
	\end{align}
	We define
	\begin{equation}
		e^{-\lambda_0}=1-\frac{2Gm_0(r)}{r},
		\qquad
		\sigma_0(r)=e^{[\nu_0(r)+\lambda_0(r)]/2}.
		\label{eq:m_sigma_def}
	\end{equation}
	The ADM mass is
	\begin{equation}
		M=\lim_{r\to\infty}m_0(r),
		\label{eq:ADM_mass}
	\end{equation}
	the conserved charge is
	\begin{equation}
		Q=8\pi\omega\int_0^\infty r^2\phi_0^2
		e^{(\lambda_0-\nu_0)/2}\,dr.
		\label{eq:Q_static}
	\end{equation}
	
	Near the origin,
	\begin{align}
		\phi_0(r)&=\phi_c-
		\frac{\phi_c}{6}
		\left(\frac{\omega^2}{\sigma_c^2}-\mu^2-2\eta\phi_c^2\right)r^2+O(r^4),
		\label{eq:phi_center_static}\\
		m_0(r)&=O(r^3),
		\label{eq:m_center_static}\\
		\sigma_0(r)&=\sigma_c+\frac{4\pi G\omega^2\phi_c^2}{\sigma_c}r^2+O(r^4),
		\label{eq:sigma_center_static}
	\end{align}
	where $\phi_c\equiv\phi_0(0)$ and $\sigma_c\equiv\sigma_0(0)$.  Asymptotic flatness and localization require
	\begin{equation}
		\phi_0\to0,
		\qquad
		\phi_0'\to0,
		\qquad
		\sigma_0\to1
		\qquad (r\to\infty).
		\label{eq:static_outer_bc}
	\end{equation}
	
	For fixed $\eta$ and fixed node number $n$, the equilibrium problem is an eigenvalue problem for $\omega$.  We write
	\begin{equation}
		M=M_n(\omega),
		\qquad
		Q=Q_n(\omega),
		\qquad
		\phi_c=\phi_{c,n}(\omega),
		\label{eq:branch_omega}
	\end{equation}
	with these relations understood parametrically near folds.  We follow the branch connected continuously to the dilute limit through its first critical point and into the adjacent segment, which is sufficient to locate the first zero crossing of the fundamental radial eigenvalue.
	
	The binding energy is
	\begin{equation}
		E_b=M-\mu Q.
		\label{eq:binding_energy}
	\end{equation}
	Let $p$ denote a regular parameter along a fixed-$n$ equilibrium
	branch at fixed self-interaction coupling. The boson star first law gives \cite{Schunck:2003kk,Herdeiro:2021mol}
	\begin{equation}
		\frac{dM}{dp}
		=
		\omega\frac{dQ}{dp}.
	\end{equation}
	Consequently, one obtains
	\begin{equation}
		\frac{dE_b}{dp}
		=
		(\omega-\mu)\frac{dQ}{dp}.
	\end{equation}
	Since $0<\omega<\mu$ for localized configurations, one obtains
	\begin{equation}
		\frac{dM}{dp}=0
		\quad\Longleftrightarrow\quad
		\frac{dQ}{dp}=0
		\quad\Longleftrightarrow\quad
		\frac{dE_b}{dp}=0.
	\end{equation}
	We therefore identify the critical point using a regular parameter along
	the equilibrium branch and quote the corresponding value of $\omega$
	as the critical frequency.
	
\begin{figure}[!htbp]
		\begin{center}
			\begin{minipage}[b]{0.237\textwidth}
				\includegraphics[width=\textwidth]{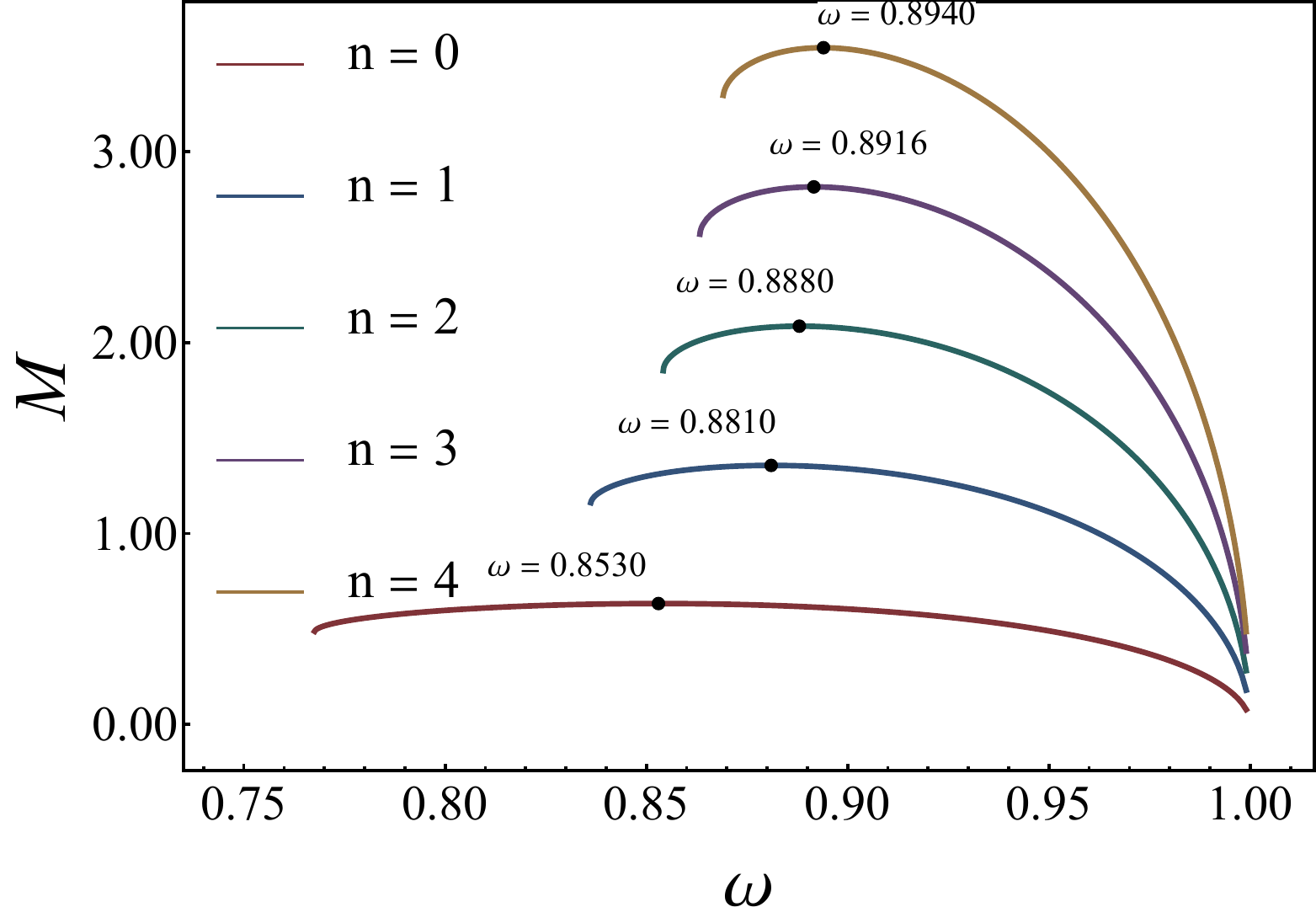}
			\end{minipage}
			\begin{minipage}[b]{0.237\textwidth}
				\includegraphics[width=\textwidth]{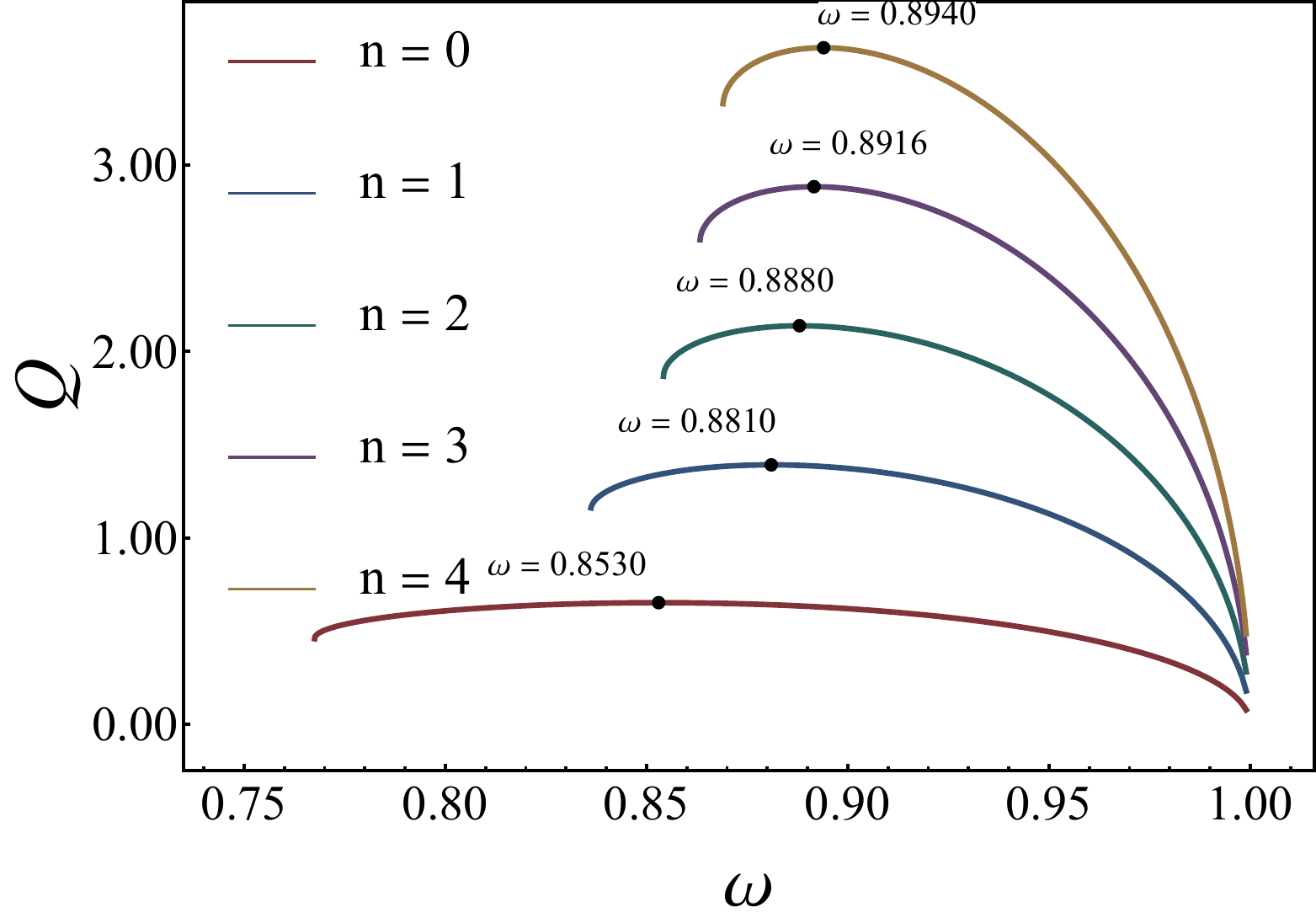}
			\end{minipage}
			\begin{minipage}[b]{0.239\textwidth}
				\includegraphics[width=\textwidth]{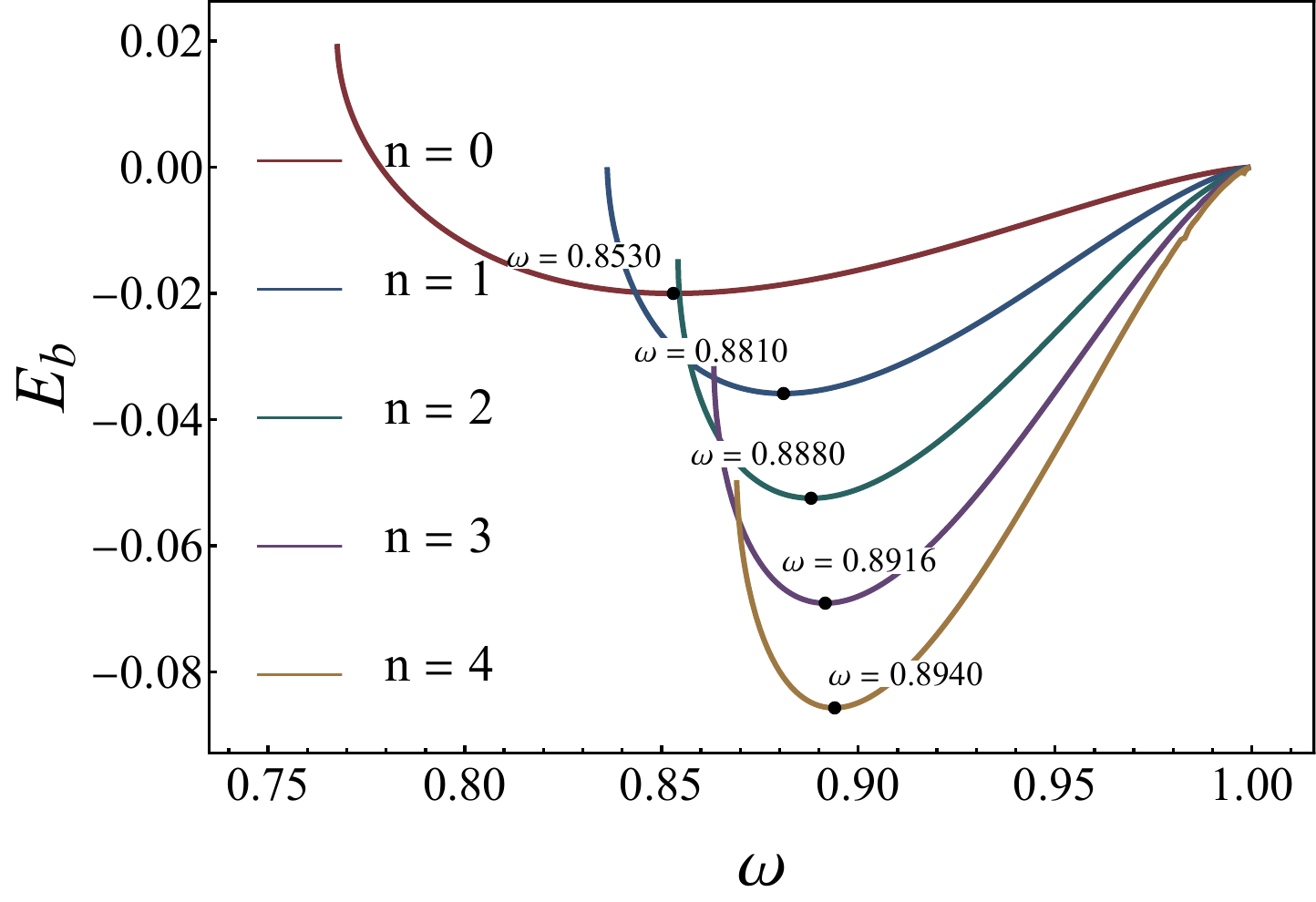}
			\end{minipage}
			\begin{minipage}[b]{0.238\textwidth}
				\includegraphics[width=\textwidth]{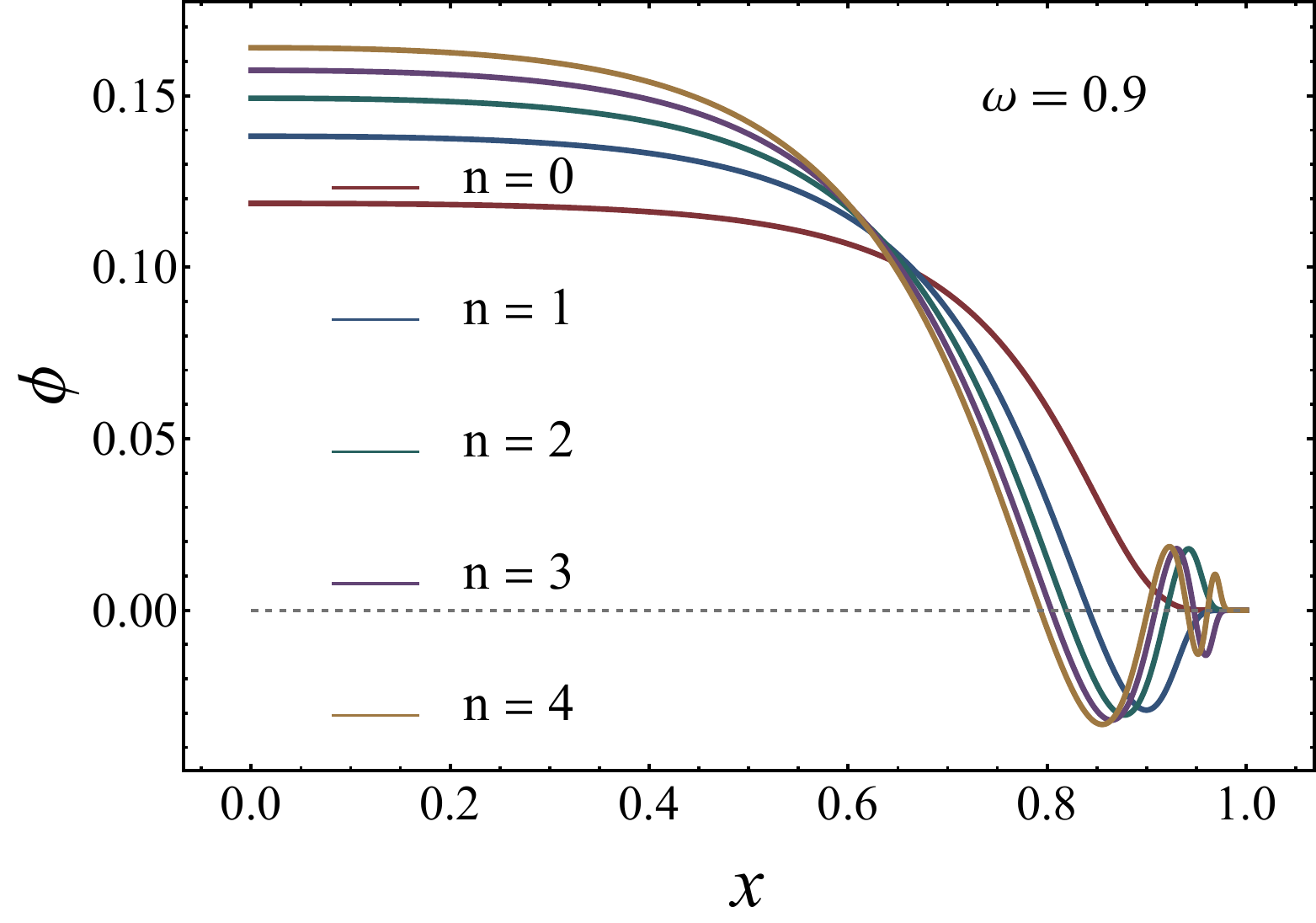}
			\end{minipage}
		\end{center}
		\caption{Background properties of boson stars with $\eta=0$ and
			$n=0,\ldots,4$. Shown are the scalar field profiles, ADM mass $M$,
			Noether charge $Q$, and binding energy $E_b$. The first extrema of
			$M$ and $Q$, together with the corresponding extremum of $E_b$, are
			marked for comparison.}
		\label{p0}
	\end{figure}

\begin{figure}[!htbp]
		\begin{center}
			\begin{minipage}[b]{0.238\textwidth}
				\includegraphics[width=\textwidth]{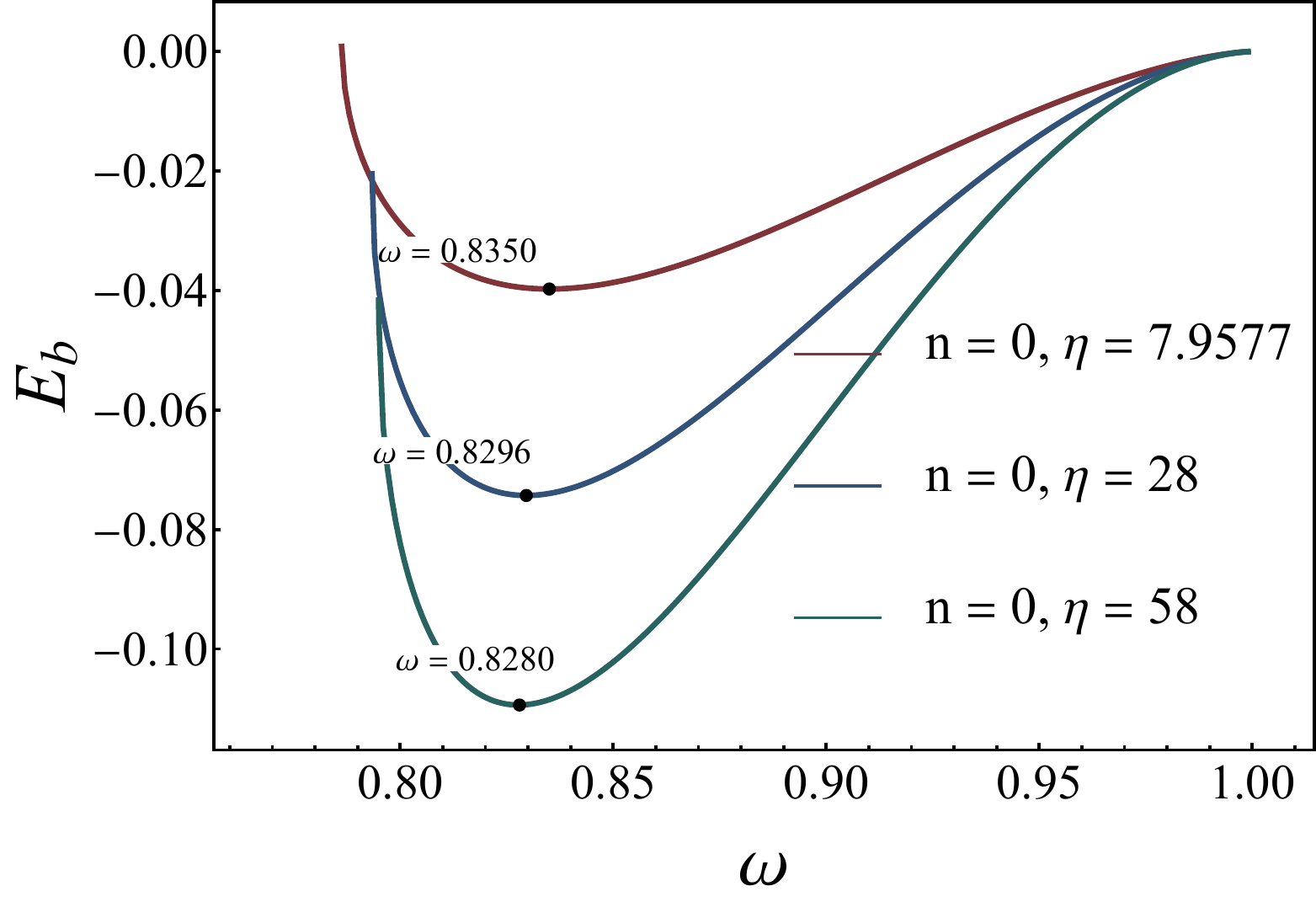}
			\end{minipage}
			\begin{minipage}[b]{0.238\textwidth}
				\includegraphics[width=\textwidth]{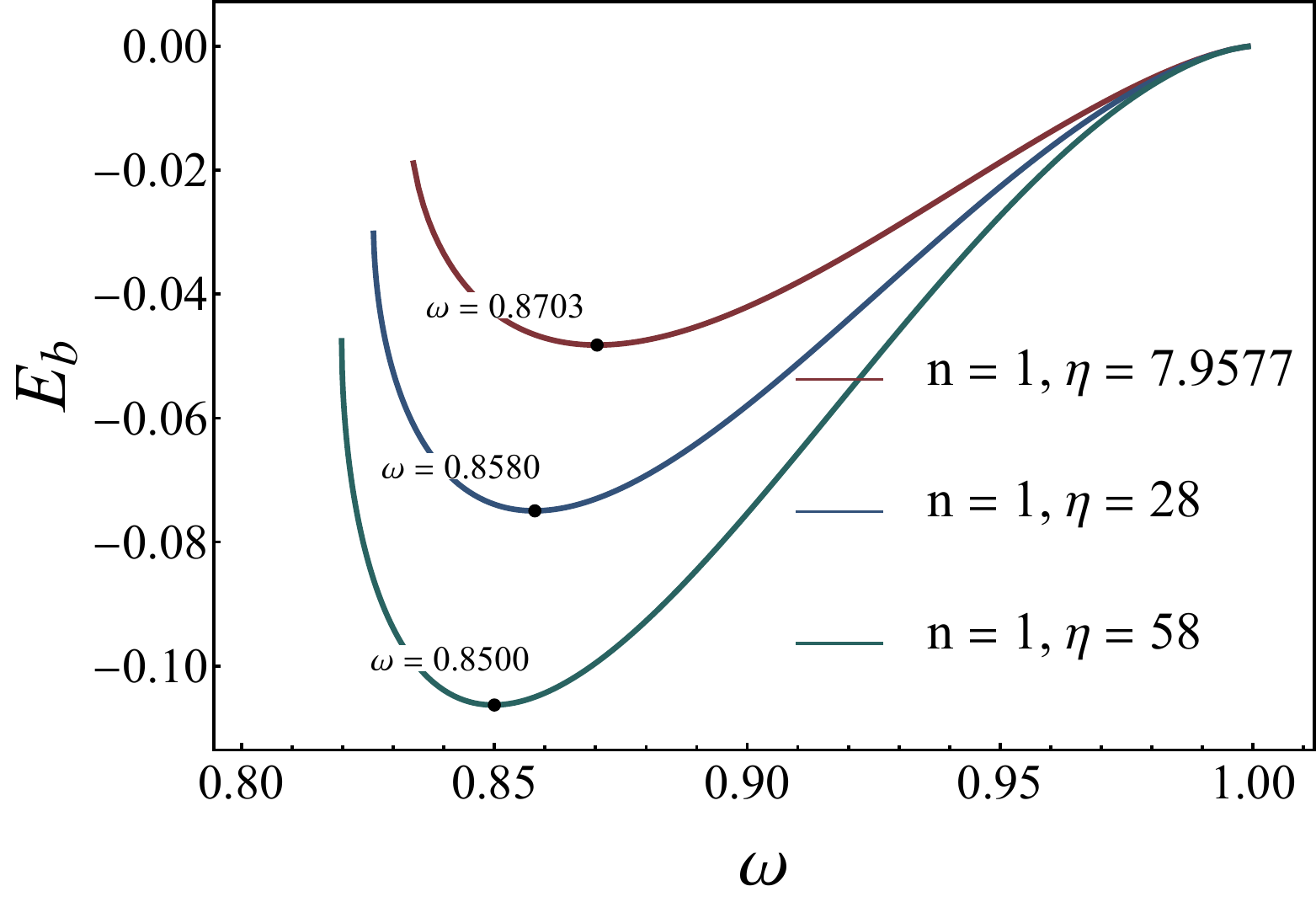}
			\end{minipage}
			\begin{minipage}[b]{0.238\textwidth}
				\includegraphics[width=\textwidth]{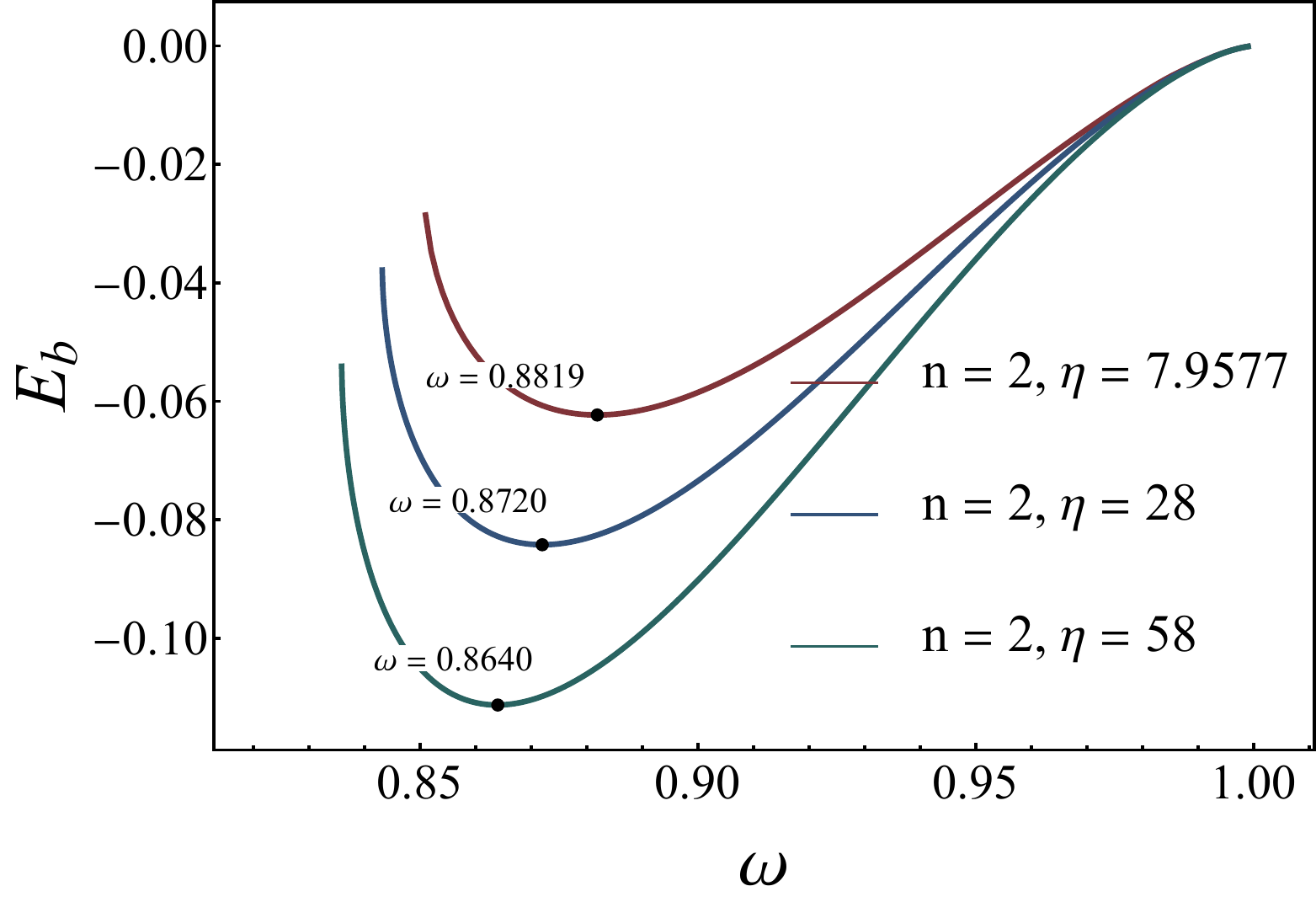}
			\end{minipage}
			\begin{minipage}[b]{0.238\textwidth}
				\includegraphics[width=\textwidth]{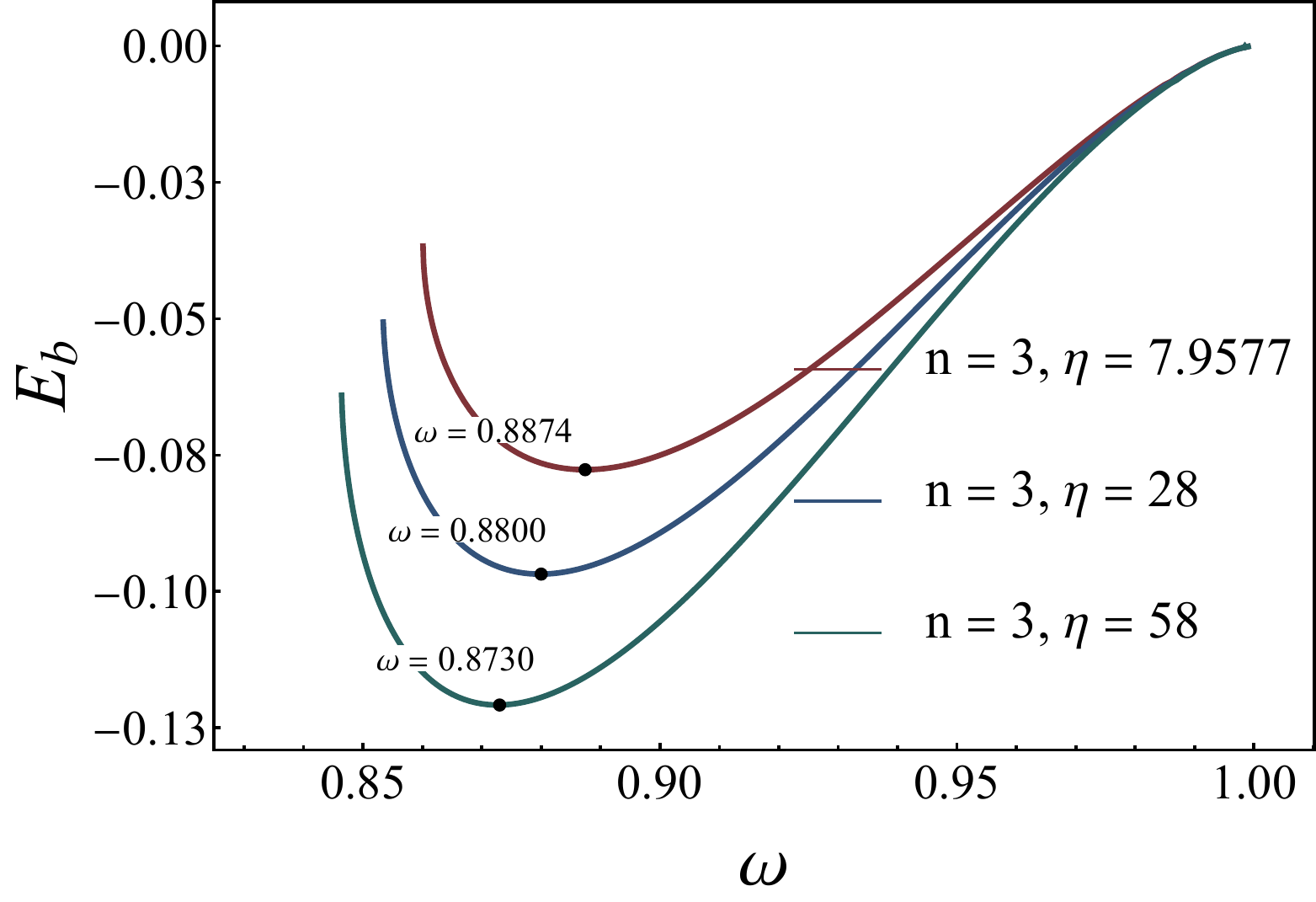}
			\end{minipage}
		\end{center}
		\caption{The binding energy $E_b$ of boson stars under three different $\eta$ for node numbers $n=0,\ldots,3$.}
		\label{p00}
	\end{figure}

	For the numerical analysis, we transform
	the range of the radial coordinate from $[0,\infty)$ to $[0,1]$: $x = \frac{r}{r+1}$. Moreover, we introduce the dimensionless quantities
	\begin{equation}
		\begin{split}
			\tilde r=\mu r,\qquad
			\tilde\omega=\frac{\omega}{\mu},\qquad
			\tilde\phi=\sqrt{4\pi G}\,\phi,\\
			\tilde M=G\mu M,\qquad
			\tilde Q=G\mu^{2}Q,
			\qquad
			\tilde\chi^{\,2}=\frac{\chi^{2}}{\mu^{2}},
		\end{split}
	\end{equation}
	together with the dimensionless self-interaction coupling
	$\tilde\eta=\eta/(4\pi G\mu^{2})$.
	Throughout the numerical calculations we set $\mu=1$ and $4\pi G=1$.
	Henceforth, we drop the tildes for notational simplicity; thus all quantities
	shown below, including $r$, $\omega$, $\phi$, $M$, $Q$, $\eta$ and $\chi^{2}$,
	refer to the corresponding rescaled dimensionless variables. For completeness, Fig.~\ref{p0} summarizes the background
	solutions for $\eta=0$ with $n=0,\ldots,4$, including the scalar field
	profiles, ADM mass $M$, Noether charge $Q$, and binding energy $E_b$.
	The first extrema of $M$ and $Q$, together with the corresponding extremum
	of $E_b$, are explicitly marked to facilitate the comparison of their
	critical frequencies. Figure~\ref{p00} shows the binding energy $E_b$ under three different $\eta$ for $n=0,\ldots,3$. Higher-node branches exhibit qualitatively similar background behavior.
	
	\section{Regular radial perturbations}
	\label{sec:perturbation}
	
	\subsection{Regular variables}
	\label{subsec:regular_variables}
	
	We decompose the scalar field as
	\begin{equation}
		\phi(t,r)=\left[\phi_1(t,r)+i\phi_2(t,r)\right]e^{-i\omega t},
		\label{eq:pert_scalar_decomp}
	\end{equation}
	with background $\phi_1=\phi_0$ and $\phi_2=0$, and write
	\begin{equation}
		\nu=\nu_0+\delta\nu,
		\qquad
		\lambda=\lambda_0+\delta\lambda.
		\label{eq:metric_pert}
	\end{equation}
	Away from the nodes of $\phi_0$, the conventional relative variables are \cite{Gleiser:1988rq,Kain:2021rmk}
	\begin{equation}
		\phi_1=\phi_0(1+\delta\phi_1),
		\qquad
		\phi_2=\phi_0\delta\phi_2.
		\label{eq:relative_variables}
	\end{equation}
	They are not suitable as numerical unknowns on excited backgrounds because their equations contain inverse powers of $\phi_0$.  Introducing
	\begin{equation}
		\dot\xi=\omega\delta\phi_2,
		\qquad
		\zeta=\xi',
		\label{eq:xi_zeta_def}
	\end{equation}
	we instead evolve the regular combinations
	\begin{equation}
		f\equiv\phi_0\delta\phi_1,
		\qquad
		g\equiv-\frac{\phi_0\xi}{\omega}.
		\label{eq:f_g_def}
	\end{equation}
	Equivalently, the perturbed scalar field can be written directly as
	\begin{equation}
		\phi(t,r)=e^{-i\omega t}\left[\phi_0(r)+f(t,r)-i\dot g(t,r)\right].
		\label{eq:direct_regular_decomp}
	\end{equation}
	We assume harmonic time dependence,
	\begin{equation}
		\{f,g,\delta\nu,\delta\lambda\}(t,r)
		=\{f,g,\mathcal N,\Lambda\}(r)e^{-i\chi t},
		\label{eq:harmonic_pert}
	\end{equation}
	where $\mathcal N\equiv\delta\nu$ and $\Lambda\equiv\delta\lambda$.  If $\chi^2<0$, we write $\chi=i\kappa$ with $\kappa=\sqrt{-\chi^2}$.
	
	The metric and enclosed-charge perturbations are
	\begin{align}
		\Lambda&=16\pi G r
		\left[\phi_0'f+\omega(\phi_0g'-\phi_0'g)\right],
		\label{eq:Lambda_regular}\\
		\delta q&=8\pi r^2e^{(\nu_0-\lambda_0)/2}
		(\phi_0g'-\phi_0'g).
		\label{eq:dq_regular}
	\end{align}
	Both expressions are regular at the nodes of $\phi_0$.
	
	\subsection{Pulsation equations}
	\label{subsec:pulsation_equations}
	
	Let
	\begin{equation}
		W_0\equiv W(\phi_0^2),
		\qquad
		W_{s0}\equiv W_s(\phi_0^2),
		\qquad
		\mathcal B\equiv\frac{2}{r}+\frac{\nu_0'-\lambda_0'}{2},
		\label{eq:W0_B_defs}
	\end{equation}
	the different self-interaction terms in the two scalar channels follow from
	\begin{equation}
		\delta\!\left[W(|\phi|^2)\phi\right]
		=\left(W_0+2\phi_0^2W_{s0}\right)\delta\phi_R
		+iW_0\delta\phi_I,
		\label{eq:potential_linearization}
	\end{equation}
	so the additional $W_{s0}$ contribution occurs only in the amplitude equation.
	
	The stress-energy perturbations are
	\begin{align}
		\delta T^t{}_t={}&
		\omega^2e^{-\nu_0}\phi_0^2\mathcal N
		+e^{-\lambda_0}{\phi_0'}^2\Lambda
		-2\omega^2e^{-\nu_0}\phi_0 f
		\nonumber\\
		&+2\omega\chi^2e^{-\nu_0}\phi_0g
		-2e^{-\lambda_0}\phi_0'f'
		-2\phi_0W_0f,
		\label{eq:dTtt_regular}\\
		\delta T^r{}_r={}&-\delta T^t{}_t-4\phi_0W_0f.
		\label{eq:dTrr_regular}
	\end{align}
	The linearized Einstein equations are found to be
	\begin{align}
		\Lambda'&=-8\pi G r e^{\lambda_0}\delta T^t{}_t
		+\lambda_0'\Lambda-\frac{\Lambda}{r},
		\label{eq:Lambda_prime}\\
		\mathcal N'&=8\pi G r e^{\lambda_0}\delta T^r{}_r
		+\nu_0'\Lambda+\frac{\Lambda}{r}.
		\label{eq:N_prime}
	\end{align}
	And the scalar equations are given by
	\begin{align}
		f''={}&-\mathcal Bf'
		-e^{\lambda_0}
		\left[(\omega^2+\chi^2)e^{-\nu_0}-W_0-2\phi_0^2W_{s0}\right]f
		\nonumber\\
		&+2\omega\chi^2e^{\lambda_0-\nu_0}g
		-\frac{\phi_0'}{2}(\mathcal N'-\Lambda')
		\nonumber\\
		&-\phi_0e^{\lambda_0}
		\left[(\omega^2e^{-\nu_0}-W_0)\Lambda
		-\omega^2e^{-\nu_0}\mathcal N\right],
		\label{eq:f_equation}\\
		g''={}&-\mathcal Bg'
		-e^{\lambda_0}
		\left[(\omega^2+\chi^2)e^{-\nu_0}-W_0\right]g
		\nonumber\\
		&+2\omega e^{\lambda_0-\nu_0}f
		+\frac{\omega}{2}\phi_0e^{\lambda_0-\nu_0}(\Lambda-\mathcal N).
		\label{eq:g_equation}
	\end{align}
	Equations~\eqref{eq:Lambda_regular}--\eqref{eq:g_equation} form a closed first-order system for $(f,f',g,g',\mathcal N)$.  They contain no inverse powers of $\phi_0$ and can therefore be integrated through the nodes of excited backgrounds.
	
	\subsection{Boundary conditions and eigenvalue problem}
	\label{subsec:numerical_method}
	
	Because the system is linear, one overall normalization is arbitrary.  Writing $A\equiv\delta\phi_1(0)$, the regular center expansion is
	\begin{align}
		\delta\phi_1(r)
		={}&A+\frac{1}{6}
		\left[
		6\zeta_1
		+A\left(
		\frac{4\omega^2-\chi^2}{\sigma_c^2}
		+4\eta\phi_c^2
		\right)
		\right]r^2
		\nonumber\\
		&+O(r^4),
		\label{eq:center_delta_phi}\\
		\zeta(r)&=\zeta_1r+O(r^3),
		\label{eq:center_zeta}\\
		\xi(r)&=\frac12\zeta_1r^2+O(r^4).
		\label{eq:xi_center}
	\end{align}
	For ordinary shooting we set $A=1$.  The remaining center value $\mathcal N(r_{\min})$ is fixed by
	\begin{align}
		\mathcal N={}&\frac{2\chi^2}{\omega^2}\xi
		+\frac{2}{\omega^2}e^{\nu_0-\lambda_0}\xi''
		+(16\pi Gr\phi_0\phi_0'+4)\delta\phi_1
		\nonumber\\
		&+\frac{2}{\omega^2}e^{\nu_0-\lambda_0}
		\left(\frac2r+2\frac{\phi_0'}{\phi_0}-\lambda_0'
		+8\pi Gr{\phi_0'}^2\right)\xi',
		\label{eq:N_constraint}
	\end{align}
	which is evaluated at a small finite $r_{\min}$ where $\phi_0\neq0$.
	
	The two shooting parameters are $\chi^2$ and $\zeta_1$.  At large radius we require
	\begin{equation}
		f\to0,
		\qquad
		\delta q\to0,
		\qquad
		\Lambda\to0.
		\label{eq:outer_bc}
	\end{equation}
	Then one can  impose
	\begin{equation}
		\Lambda(r_{\rm out})=0,
		\qquad
		\delta q(r_{\rm out})=0,
		\label{eq:lambda_q_bc}
	\end{equation}
	and monitor $f(r_{\rm out})$ independently.  The result is accepted only after the eigenvalue and all three boundary residuals are stable over a range of $r_{\rm out}$.  We define $R_{99}$ by
	\begin{equation}
		m_0(R_{99})=0.99M,
		\label{eq:R99_def}
	\end{equation}
	and quote results from an $r_{\rm out}/R_{99}$ plateau rather than from a single outer boundary.
	
	Since the present work concerns only the fundamental radial mode, no overtone classification is required.  For each fixed $n$, we first locate the lowest real root on a reference background by scanning the boundary residual over a finite interval in $\chi^2$.  The branch $\chi_{0,n}^2(\omega)$ is then followed continuously, using the neighboring solution as the initial guess; selected anchor points are rescanned to verify that no lower real root has been missed.  The sign of this lowest eigenvalue determines stability within the spherically symmetric perturbation sector:
	\begin{align}
		\chi_{0,n}^2>0
		&\quad\Rightarrow\quad
		\text{oscillatory fundamental mode},
		\nonumber\\
		\chi_{0,n}^2<0
		&\quad\Rightarrow\quad
		\text{radial instability}.
		\label{eq:stability_criterion}
	\end{align}
	
	\subsection{Zero-mode diagnostic}
	\label{subsec:zero_mode_diagnostic}
	
	Near a turning point, direct shooting for a very small $\chi^2$ may be ill-conditioned.  At fixed background and fixed $\chi^2$, the regular center data are spanned by the two solutions
	\begin{equation}
		(A,\zeta_1)=(1,0),
		\qquad
		(A,\zeta_1)=(0,1),
		\label{eq:zero_basis}
	\end{equation}
	which we denote by $y_A$ and $y_\zeta$.  Their outer-boundary residuals define
	\begin{equation}
		\mathcal R[y](r_{\rm out})=
		\begin{pmatrix}
			\Lambda(r_{\rm out})\\
			\delta q(r_{\rm out})\\
			f(r_{\rm out})
		\end{pmatrix},
		\label{eq:zero_residual_vector}
	\end{equation}
	and the $3\times2$ boundary map
	\begin{equation}
		\mathcal C(\chi^2;r_{\rm out})=
		\left(
		\mathcal R[y_A]\quad\mathcal R[y_\zeta]
		\right).
		\label{eq:zero_matrix}
	\end{equation}
	An exact zero mode at $\chi^2=0$ requires
	\begin{equation}
		\operatorname{rank}\mathcal C(0;r_{\rm out})<2.
		\label{eq:zero_rank}
	\end{equation}
	To reduce the sensitivity to the different numerical scales of the
	three boundary residuals, we normalize the rows of $\mathcal C$ and
	monitor the singular value ratio
	\begin{equation}
		\epsilon(\chi^2;r_{\rm out})
		=\frac{s_2}{s_1},
		\qquad s_1\ge s_2\ge0,
		\label{eq:svd_ratio}
	\end{equation}
	is evaluated.  A small value of $\epsilon(0;r_{\rm out})$ indicates compatibility with a zero mode, but by itself cannot distinguish an exact zero from a nearby low-frequency mode.  We therefore also scan $\epsilon(\chi^2;r_{\rm out})$ in a neighborhood of zero and require its minimum to occur at $\chi^2=0$ within the numerical uncertainty.  The critical point is finally confirmed by a sign change of the directly computed branch $\chi_{0,n}^2(\omega)$ and by stability under changes of $r_{\rm out}$.
	
	\section{Results}
	\label{sec:results}
	
	\subsection{Fundamental radial modes}
	
	For excited boson stars, the behavior of the fundamental radial eigenvalue along fixed-node branches has not been systematically established. We therefore compute the lowest radial eigenvalue $\chi_{0,n}^{2}$ for the branches with $n=0,1,2,3$, considering several representative values of the quartic self-interaction coupling $\eta$.
	
\begin{figure}[H]
		\begin{center}
			\begin{minipage}[b]{0.238\textwidth}
				\includegraphics[width=\textwidth]{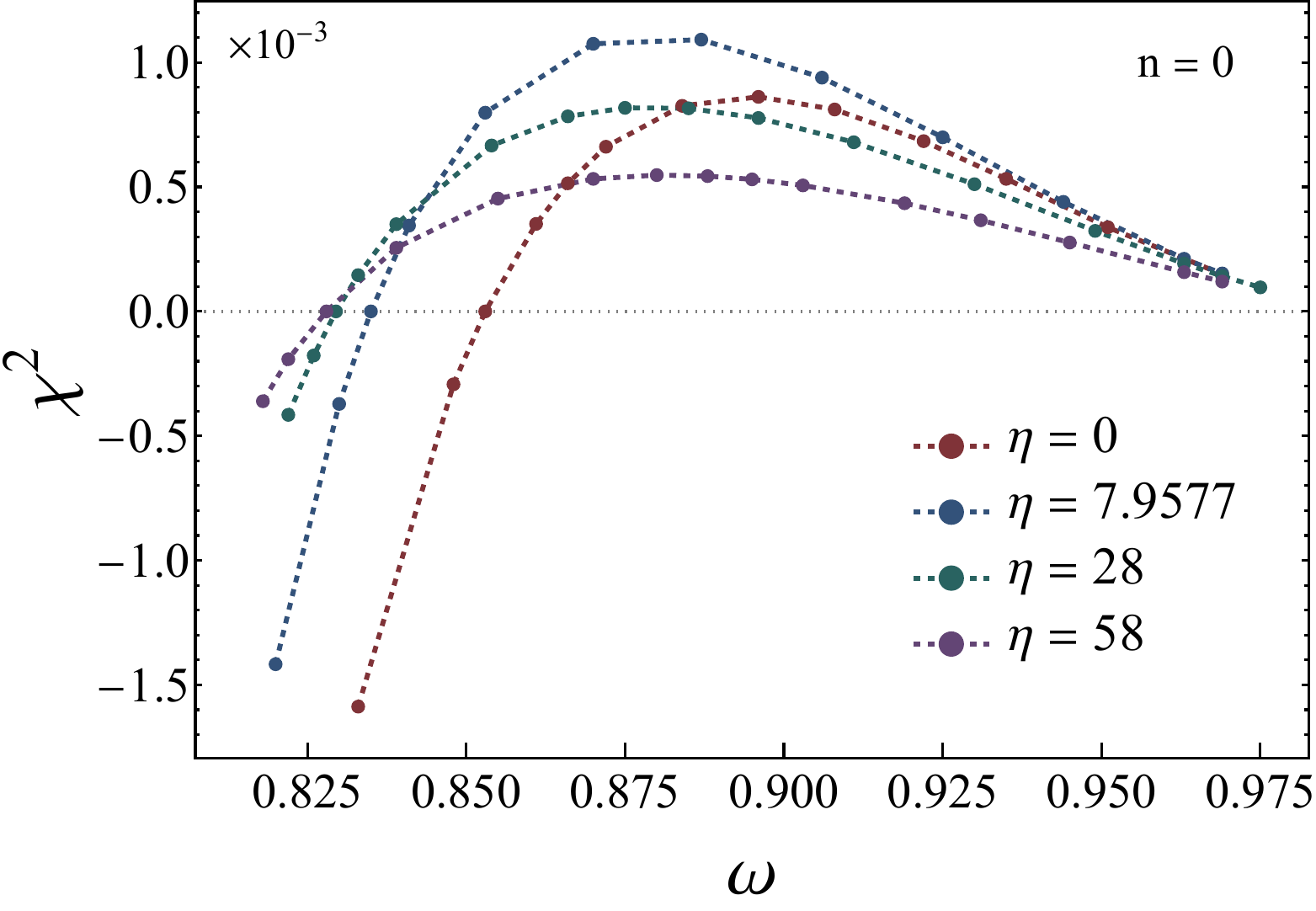}
			\end{minipage}
			\begin{minipage}[b]{0.238\textwidth}
				\includegraphics[width=\textwidth]{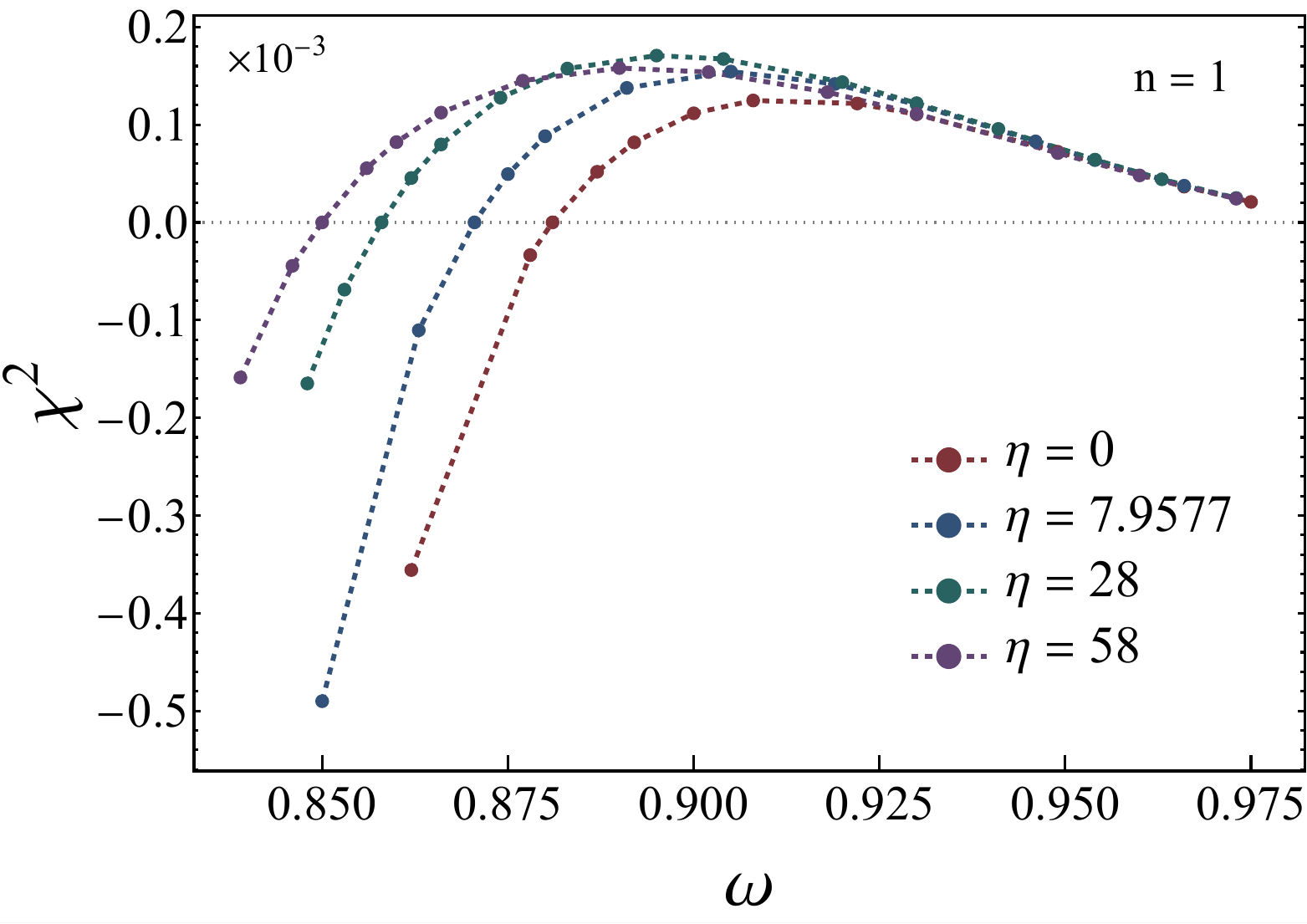}
			\end{minipage}
			\begin{minipage}[b]{0.238\textwidth}
				\includegraphics[width=\textwidth]{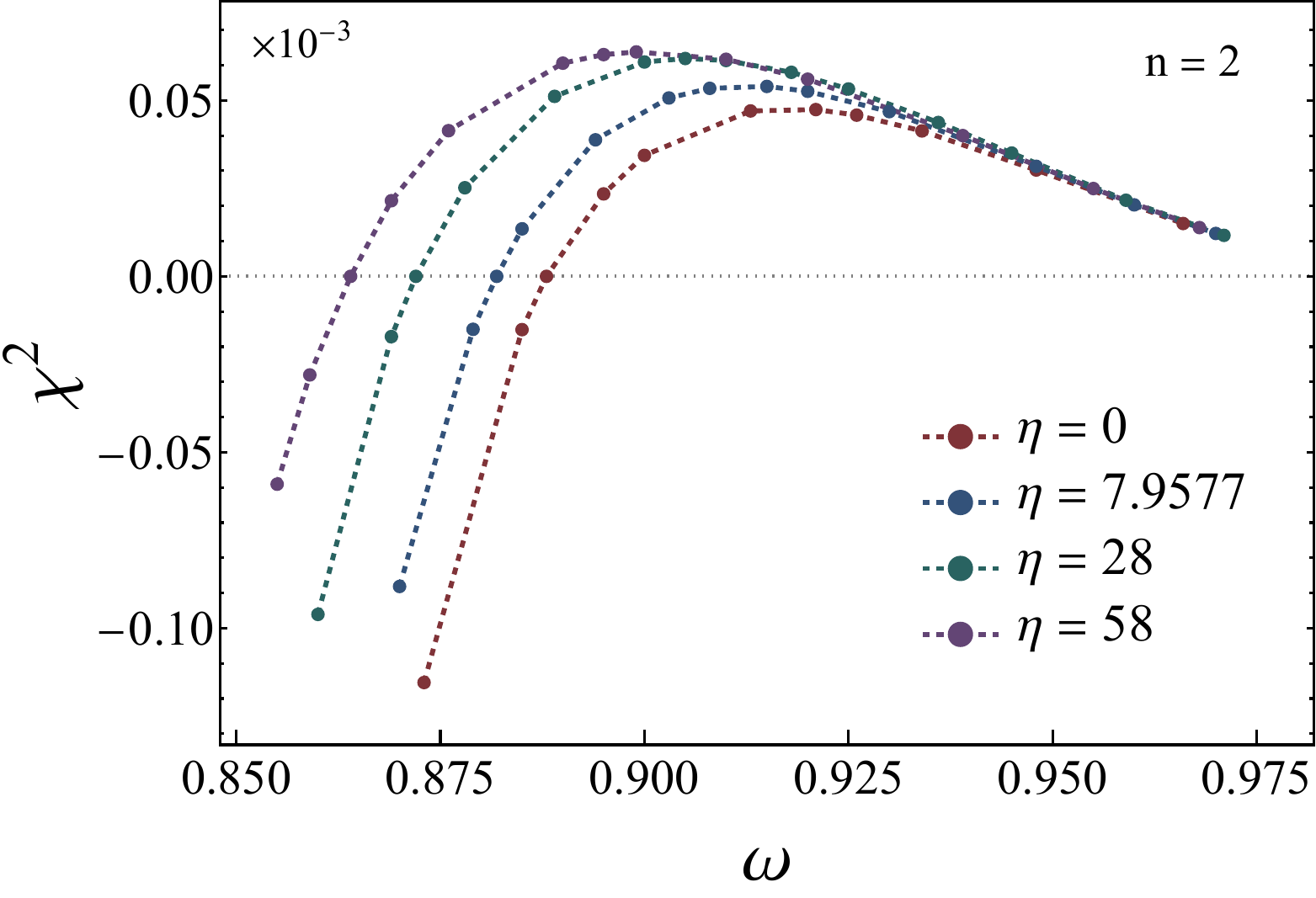}
			\end{minipage}
			\begin{minipage}[b]{0.238\textwidth}
				\includegraphics[width=\textwidth]{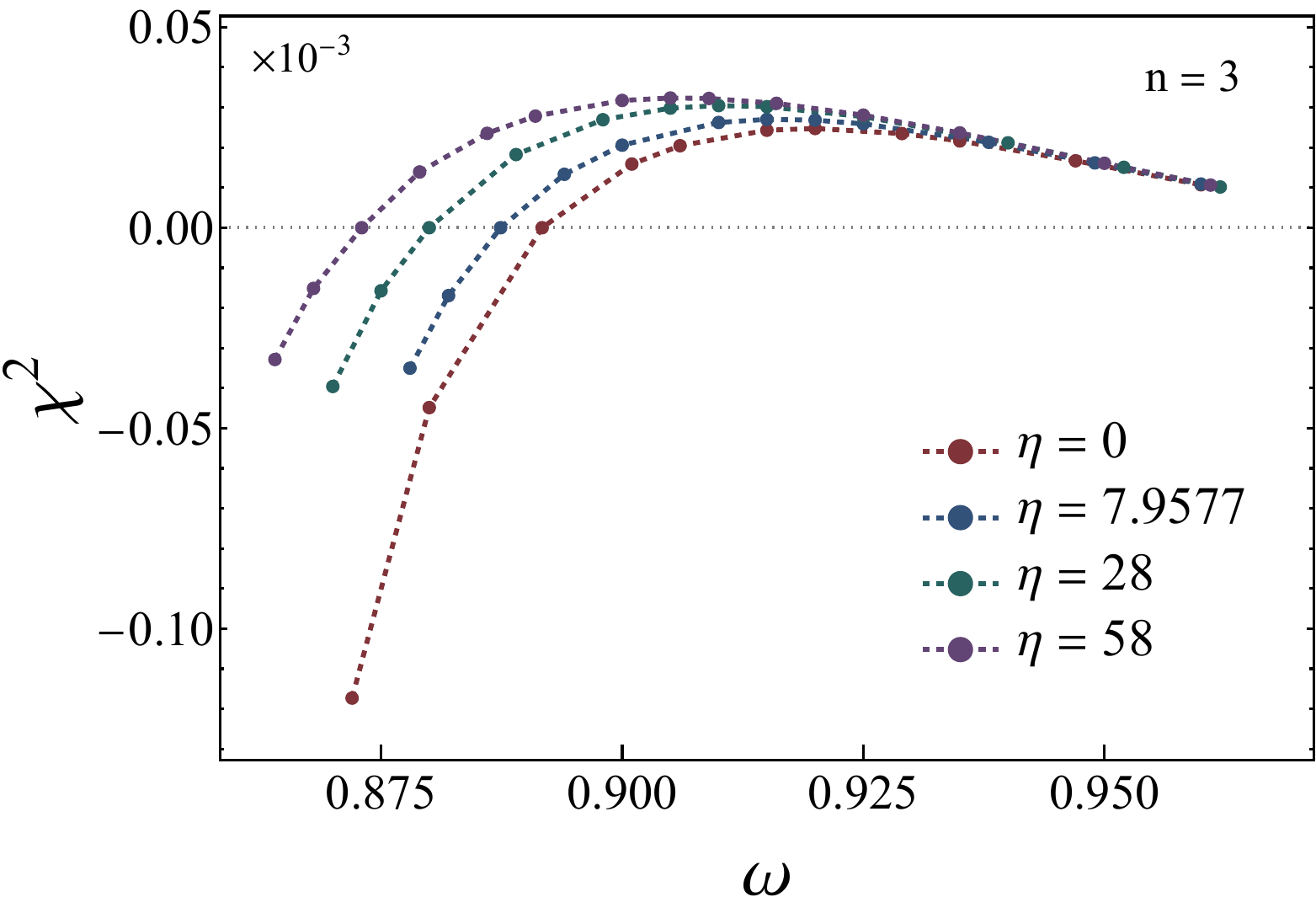}
			\end{minipage}
		\end{center}
		\caption{Fundamental radial eigenvalue $\chi_{0,n}^{2}$ for the fixed node branches $n=0,1,2,3$ and several values of the quartic coupling $\eta$. Dashed segments are guides to the eye and do not represent additional computed data.}
		\label{p1}
	\end{figure}

	Figure~\ref{p1} shows $\chi_{0,n}^{2}$ as a function of the background frequency $\omega$ for $\eta=0$, $7.9577$, $28$, and $58$. For all cases considered, the fundamental eigenvalue changes sign along the branch, separating a region with $\chi_{0,n}^{2}<0$ from one with $\chi_{0,n}^{2}>0$. Thus, each fixed-node branch exhibits a radial zero mode that marks the boundary between radially unstable and oscillatory configurations within the spherically symmetric perturbation sector. Several general features can be read from the numerical results. As the node number $n$ increases, the characteristic scale of $\chi_{0,n}^{2}$ becomes progressively smaller, indicating a systematic softening of the fundamental radial mode for higher excited states. For a fixed $n$, increasing the self-interaction coupling shifts the zero-mode crossing toward lower values of $\omega$. We further find that, for the excited boson stars considered here, the zero crossing of the fundamental radial mode is in one-to-one correspondence with the first critical point of the equilibrium branch, as identified consistently from the extrema of $M$, $Q$, and $E_b$. This correspondence holds both for mini-boson stars and in the presence of quartic self-interaction. Detailed numerical comparisons of the corresponding critical frequencies are given in Tables~\ref{tab:radial-n0}--\ref{tab:radial-n3}. We have extended the calculation to higher excited states up to $n=7$ and found the same behavior in all cases examined; these results are therefore not displayed separately.
	
	\subsection{Dynamical information encoded in the radial mode}
	
	Linear radial perturbations and nonlinear time evolutions probe different aspects of boson star stability. The former determine the spectrum of infinitesimal radial oscillations around an equilibrium configuration, whereas the latter follow the full dynamical response over a finite time interval. It is therefore not obvious a priori whether stability thresholds inferred from time evolutions should leave a simple quantitative signature in the radial spectrum. Our results nevertheless indicate that such a connection exists.
	In \cite{Brito:2023fwr}, spherical evolutions of boson stars were performed for different node numbers and self-interaction couplings. For fixed $\omega=0.90$ and $0.92$, the authors identified the first sampled coupling for which the configurations remained stable over the simulated time interval. We evaluate the fundamental radial eigenvalue $\chi^2_{0,n}$ for these reported threshold models. The results for $n=1,\ldots,7$, summarized in Table~\ref{tab:dynamical_threshold}, reveal a clear empirical correlation between $\chi^2_{0,n}$, $n$, and $\eta$.
	
	\begin{table}[htbp!]
		\centering
		\caption{Fundamental radial eigenvalue $\chi_{0,n}^{2}$ evaluated at the dynamical stability thresholds.}
		\label{tab:dynamical_threshold}
		\begin{tabular}{c|cc|cc}
			\hline\hline
			$n$
			& $\eta\;(\omega=0.90)$
			& $\chi_{0,n}^{2}$
			& $\eta\;(\omega=0.92)$
			& $\chi_{0,n}^{2}$ \\
			\hline
			1 & 75  & $1.4556\times10^{-4}$ & 74  & $1.2233\times10^{-4}$ \\
			2 & 160 & $5.5615\times10^{-5}$ & 150 & $4.8097\times10^{-5}$ \\
			3 & 300 & $2.8242\times10^{-5}$ & 250 & $2.5312\times10^{-5}$ \\
			4 & 450 & $1.7279\times10^{-5}$ & 390 & $1.5378\times10^{-5}$ \\
			5 & 650 & $1.1322\times10^{-5}$ & 550 & $1.0034\times10^{-5}$ \\
			6 & 850 & $8.3001\times10^{-6}$ & 760 & $6.8388\times10^{-6}$ \\
			7 & 1160 & $5.8601\times10^{-6}$ & 1000 & $4.9816\times10^{-6}$ \\
			\hline\hline
		\end{tabular}
	\end{table}

	Using only the first four excited states, $n=1,\ldots,4$, we find that the threshold data can be accurately described by the empirical relations
	\begin{equation}
		\frac{\chi_{0,n}^{2}}{\mu^{2}}
		\simeq
		\frac{1.07991\times10^{-2}}
		{\eta\,n^{0.237199}},
		\qquad
		\frac{\omega}{\mu}=0.90 ,
		\label{eq:fit090}
	\end{equation}
	\begin{equation}
		\frac{\chi_{0,n}^{2}}{\mu^{2}}
		\simeq
		\frac{8.98299\times10^{-3}}
		{\eta\,n^{0.303924}},
		\qquad
		\frac{\omega}{\mu}=0.92 ,
		\label{eq:fit092}
	\end{equation}
	with $R^{2}\simeq0.99947$ and $0.99981$, respectively. The $n=5, 6, 7$ configurations were not used in obtaining these fits and therefore
	provide an independent test of the scaling. For these higher excited
	states, the relative deviations between the fitted and directly
	calculated eigenvalues remain below $0.3\%$, showing that the same
	scaling continues beyond the data used to determine it.
	
	The decreasing magnitude of $\chi^2_{0,n}$ indicates that the fundamental radial mode becomes progressively softer for higher excited states. The fact that the nonlinear threshold models exhibit the same spectral trend suggests that this softening is not merely a property of the equilibrium sequence, but is also correlated with the self-interaction strength required to sustain excited configurations during nonlinear evolution. Once calibrated from a small set of configurations, the empirical relation may therefore be used to estimate the self-interaction strength at which dynamically stable configurations are expected to emerge. Although such an estimate cannot replace nonlinear time evolutions, it can provide a useful guide to the relevant region of parameter space before more computationally expensive evolutions are performed.
	
	\section{Conclusions}
	\label{sec:conclusions}

	We have studied the lowest radial mode of ground state and excited boson stars, including both the mini-boson star limit and quartic self-interactions. The main technical difficulty for excited backgrounds arises from radial nodes in the equilibrium scalar profile, where conventional relative perturbation variables become singular. By introducing additive variables regular at these nodes, we formulate a radial eigenvalue problem that can be integrated continuously across nodeful backgrounds. The nodeless results benchmark the method, while the same formulation allows the fundamental mode to be systematically followed along higher-node branches. For all configurations examined, the first zero of the constrained fundamental radial eigenvalue coincides, within numerical resolution, with the first critical point of the equilibrium branch, independently identified from the ADM mass, the Noether charge, and the binding energy. This correspondence holds for both mini boson stars and quartically self-interacting configurations.
	
	We have also compared the radial spectrum with threshold models reported in nonlinear spherical evolutions of self-interacting excited boson stars. The corresponding eigenvalues show a simple empirical dependence on node number and self-interaction strength, and the scaling obtained from the first few excited states continues to describe the higher-node models considered here. Since the nonlinear thresholds were obtained from finite resolution scans in coupling space and over finite evolution times, this relation should presently be regarded as a phenomenological correlation rather than an exact stability criterion. More general radial and nonradial perturbation analyses would provide a direct test of whether this spectral correlation can become a genuinely predictive diagnostic of boson star dynamics.

	\section*{Acknowledgements}
	This work was supported by the National Natural Science Foundation of China (Grants No. 12405055, No. 12205129, No. 12475051,No. 12375051,and No. 12421005
	), the Key Project of the Department of Education of Hunan Province(No. 25A0084); the science and technology innovation Program of Hunan Province under grant No. 2024RC1050; the Natural Science Foundation of Hunan Province under grant  No. 2023JJ30384; the innovative research group of Hunan Province under grant No. 2024JJ1006. 
	
	\appendix
	\label{app1}

	\begingroup
	\small
	\renewcommand{\arraystretch}{1.12}
	\setlength{\tabcolsep}{17pt}
	\setlength{\LTleft}{\fill}
	\setlength{\LTright}{\fill}
	
	
	\begin{longtable*}{@{}cccccc@{}}
		\caption{Radial perturbation data for the ground-state solutions $(n=0)$.}
		\label{tab:radial-n0}\\
		\hline\hline
		$n$ & $\eta$ & $\phi(0)$ & $\omega$ & $\chi^2$ & $\zeta$ \\
		\hline\hline
		\endfirsthead
		\multicolumn{6}{c}{\tablename~\thetable\ (continued)}\\
		\hline\hline
		$n$ & $\eta$ & $\phi(0)$ & $\omega$ & $\chi^2$ & $\zeta$ \\
		\hline\hline
		\endhead
		\hline
		\multicolumn{6}{r}{Continued on next page}\\
		\endfoot
		\hline\hline
		\endlastfoot
		
		$0$ & $0$ & $0.03300$ & $0.969$ & $1.4915\times10^{-4}$ & $-0.72315$ \\
		$0$ & $0$ & $0.05353$ & $0.951$ & $3.3796\times10^{-4}$ & $-0.7629$ \\
		$0$ & $0$ & $0.07274$ & $0.935$ & $5.3236\times10^{-4}$ & $-0.8038$ \\
		$0$ & $0$ & $0.08911$ & $0.922$ & $6.8390\times10^{-4}$ & $-0.84177$ \\
		$0$ & $0$ & $0.10760$ & $0.908$ & $8.1108\times10^{-4}$ & $-0.88848$ \\
		$0$ & $0$ & $0.12424$ & $0.896$ & $8.6184\times10^{-4}$ & $-0.93439$ \\
		$0$ & $0$ & $0.14173$ & $0.884$ & $8.2569\times10^{-4}$ & $-0.98698$ \\
		$0$ & $0$ & $0.16017$ & $0.872$ & $6.6141\times10^{-4}$ & $-1.0479$ \\
		$0$ & $0$ & $0.16981$ & $0.866$ & $5.1470\times10^{-4}$ & $-1.0821$ \\
		$0$ & $0$ & $0.17806$ & $0.861$ & $3.5152\times10^{-4}$ & $-1.1129$ \\
		$0$ & $0$ & $0.19175$ & $0.853$ & $0$ & -1.1670 \\
		$0$ & $0$ & $0.20064$ & $0.848$ & $-2.9293\times10^{-4}$ & $-1.2044$ \\
		$0$ & $0$ & $0.22909$ & $0.833$ & $-1.5875\times10^{-3}$ & $-1.3373$ \\
		\hline
		
		$0$ & $7.9577$ & $0.03026$ & $0.969$ & $1.5231\times10^{-4}$ & $-0.72568$ \\
		$0$ & $7.9577$ & $0.03587$ & $0.963$ & $2.1140\times10^{-4}$ & $-0.73932$ \\
		$0$ & $7.9577$ & $0.05339$ & $0.944$ & $4.3991\times10^{-4}$ & $-0.78817$ \\
		$0$ & $7.9577$ & $0.07079$ & $0.925$ & $6.9931\times10^{-4}$ & $-0.84713$ \\
		$0$ & $7.9577$ & $0.08846$ & $0.906$ & $9.3906\times10^{-4}$ & $-0.91902$ \\
		$0$ & $7.9577$ & $0.10685$ & $0.887$ & $1.0917\times10^{-3}$ & $-1.0082$ \\
		$0$ & $7.9577$ & $0.14353$ & $0.853$ & $7.9819\times10^{-4}$ & $-1.237$ \\
		$0$ & $7.9577$ & $0.15854$ & $0.841$ & $3.4484\times10^{-4}$ & $-1.3528$ \\
		$0$ & $7.9577$ & $0.16669$ & $0.835$ & $0$ & -1.4218 \\
		$0$ & $7.9577$ & $0.17390$ & $0.830$ & $-3.7162\times10^{-4}$ & $-1.4866$ \\
		$0$ & $7.9577$ & $0.18983$ & $0.820$ & $-1.4172\times10^{-3}$ & $-1.6429$ \\
		\hline
		
		$0$ & $28$ & $0.02110$ & $0.975$ & $9.6521\times10^{-5}$ & $-0.71562$ \\
		$0$ & $28$ & $0.02527$ & $0.969$ & $1.4161\times10^{-4}$ & $-0.72962$ \\
		$0$ & $28$ & $0.02923$ & $0.963$ & $1.9207\times10^{-4}$ & $-0.74489$ \\
		$0$ & $28$ & $0.03788$ & $0.949$ & $3.2342\times10^{-4}$ & $-0.78263$ \\
		$0$ & $28$ & $0.04883$ & $0.930$ & $5.1099\times10^{-4}$ & $-0.84297$ \\
		$0$ & $28$ & $0.05946$ & $0.911$ & $6.7950\times10^{-4}$ & $-0.91545$ \\
		$0$ & $28$ & $0.08050$ & $0.875$ & $8.1771\times10^{-4}$ & $-1.1023$ \\
		$0$ & $28$ & $0.08629$ & $0.866$ & $7.8348\times10^{-4}$ & $-1.1643$ \\
		$0$ & $28$ & $0.09458$ & $0.854$ & $6.6639\times10^{-4}$ & $-1.2614$ \\
		$0$ & $28$ & $0.10633$ & $0.839$ & $3.5037\times10^{-4}$ & $-1.4161$ \\
		$0$ & $28$ & $0.11164$ & $0.833$ & $1.4560\times10^{-4}$ & $-1.4931$ \\
		$0$ & $28$ & $0.11497$ & $0.8296$ & $0$ & $-1.5434$ \\
		$0$ & $28$ & $0.11849$ & $0.826$ & $-1.7714\times10^{-4}$ & $-1.5986$ \\
		$0$ & $28$ & $0.12281$ & $0.822$ & $-4.1554\times10^{-4}$ & $-1.669$ \\
		\hline
		
		$0$ & $58$ & $0.02072$ & $0.969$ & $1.1975\times10^{-4}$ & $-0.7325$ \\
		$0$ & $58$ & $0.02354$ & $0.963$ & $1.5756\times10^{-4}$ & $-0.7479$ \\
		$0$ & $58$ & $0.03119$ & $0.945$ & $2.7668\times10^{-4}$ & $-0.79904$ \\
		$0$ & $58$ & $0.03669$ & $0.931$ & $3.6613\times10^{-4}$ & $-0.84464$ \\
		$0$ & $58$ & $0.04130$ & $0.919$ & $4.3454\times10^{-4}$ & $-0.88861$ \\
		$0$ & $58$ & $0.05683$ & $0.880$ & $5.4761\times10^{-4}$ & $-1.0763$ \\
		$0$ & $58$ & $0.06123$ & $0.870$ & $5.3239\times10^{-4}$ & $-1.1406$ \\
		$0$ & $58$ & $0.06842$ & $0.855$ & $4.5314\times10^{-4}$ & $-1.257$ \\
		$0$ & $58$ & $0.07733$ & $0.839$ & $2.5548\times10^{-4}$ & $-1.4202$ \\
		$0$ & $58$ & $0.08466$ & $0.828$ & $0$ & $-1.5708$ \\
		$0$ & $58$ & $0.08931$ & $0.822$ & $-1.9178\times10^{-4}$ & $-1.6742$ \\
		$0$ & $58$ & $0.09278$ & $0.818$ & $-3.6091\times10^{-4}$ & $-1.7552$ \\
	\end{longtable*}
	
	\clearpage
	
	
	\begin{longtable*}{@{}cccccc@{}}
		\caption{Radial perturbation data for the first excited-state solutions ($n=1$).}
		\label{tab:radial-n1}\\
		\hline\hline
		$n$ & $\eta$ & $\phi(0)$ & $\omega$ & $\chi^2$ & $\zeta$ \\
		\hline\hline
		\endfirsthead
		\multicolumn{6}{c}{\tablename~\thetable\ (continued)}\\
		\hline\hline
		$n$ & $\eta$ & $\phi(0)$ & $\omega$ & $\chi^2$ & $\zeta$ \\
		\hline\hline
		\endhead
		\hline
		\multicolumn{6}{r}{Continued on next page}\\
		\endfoot
		\hline\hline
		\endlastfoot
		
		$1$ & $0$ & $0.02858$ & $0.975$ & $2.0937\times10^{-5}$ & $-0.73082$ \\
		$1$ & $0$ & $0.03958$ & $0.966$ & $3.6609\times10^{-5}$ & $-0.75865$ \\
		$1$ & $0$ & $0.06160$ & $0.949$ & $7.2285\times10^{-5}$ & $-0.8203$ \\
		$1$ & $0$ & $0.08855$ & $0.930$ & $1.1063\times10^{-4}$ & $-0.90807$ \\
		$1$ & $0$ & $0.10080$ & $0.922$ & $1.2171\times10^{-4}$ & $-0.95312$ \\
		$1$ & $0$ & $0.12387$ & $0.908$ & $1.2479\times10^{-4}$ & $-1.0479$ \\
		$1$ & $0$ & $0.13819$ & $0.900$ & $1.1166\times10^{-4}$ & $-1.1141$ \\
		$1$ & $0$ & $0.15353$ & $0.892$ & $8.1861\times10^{-5}$ & $-1.192$ \\
		$1$ & $0$ & $0.16373$ & $0.887$ & $5.1778\times10^{-5}$ & $-1.2483$ \\
		$1$ & $0$ & $0.17674$ & $0.881$ & $0$ & $-1.3255$ \\
		$1$ & $0$ & $0.18360$ & $0.878$ & $-3.3494\times10^{-5}$ & $-1.369$ \\
		$1$ & $0$ & $0.22581$ & $0.862$ & $-3.5588\times10^{-4}$ & $-1.6832$ \\
		\hline
		
		$1$ & $7.9577$ & $0.02899$ & $0.973$ & $2.4594\times10^{-5}$ & $-0.73741$ \\
		$1$ & $7.9577$ & $0.03644$ & $0.966$ & $3.7543\times10^{-5}$ & $-0.7596$ \\
		$1$ & $7.9577$ & $0.04926$ & $0.954$ & $6.3768\times10^{-5}$ & $-0.80222$ \\
		$1$ & $7.9577$ & $0.05790$ & $0.946$ & $8.2914\times10^{-5}$ & $-0.83434$ \\
		$1$ & $7.9577$ & $0.07560$ & $0.930$ & $1.2063\times10^{-4}$ & $-0.90963$ \\
		$1$ & $7.9577$ & $0.08827$ & $0.919$ & $1.4169\times10^{-4}$ & $-0.97212$ \\
		$1$ & $7.9577$ & $0.10529$ & $0.905$ & $1.5440\times10^{-4}$ & $-1.0688$ \\
		$1$ & $7.9577$ & $0.12382$ & $0.891$ & $1.3764\times10^{-4}$ & $-1.1923$ \\
		$1$ & $7.9577$ & $0.13997$ & $0.880$ & $8.8216\times10^{-5}$ & $-1.3172$ \\
		$1$ & $7.9577$ & $0.14795$ & $0.875$ & $4.9455\times10^{-5}$ & $-1.3855$ \\
		$1$ & $7.9577$ & $0.15558$ & $0.8703$ & $0$ & $-1.4554$ \\
		$1$ & $7.9577$ & $0.16949$ & $0.863$ & $-1.1039\times10^{-4}$ & $-1.5935$ \\
		$1$ & $7.9577$ & $0.19932$ & $0.850$ & $-4.9014\times10^{-4}$ & $-1.9455$ \\
		\hline
		
		$1$ & $28$ & $0.02504$ & $0.973$ & $2.4897\times10^{-5}$ & $-0.73854$ \\
		$1$ & $28$ & $0.03285$ & $0.963$ & $4.4108\times10^{-5}$ & $-0.77127$ \\
		$1$ & $28$ & $0.03951$ & $0.954$ & $6.4221\times10^{-5}$ & $-0.80395$ \\
		$1$ & $28$ & $0.04875$ & $0.941$ & $9.5658\times10^{-5}$ & $-0.8573$ \\
		$1$ & $28$ & $0.05640$ & $0.930$ & $1.2204\times10^{-4}$ & $-0.90912$ \\
		$1$ & $28$ & $0.06337$ & $0.920$ & $1.4359\times10^{-4}$ & $-0.96265$ \\
		$1$ & $28$ & $0.08161$ & $0.895$ & $1.7073\times10^{-4}$ & $-1.1334$ \\
		$1$ & $28$ & $0.09127$ & $0.883$ & $1.5746\times10^{-4}$ & $-1.2427$ \\
		$1$ & $28$ & $0.09919$ & $0.874$ & $1.2765\times10^{-4}$ & $-1.3427$ \\
		$1$ & $28$ & $0.10694$ & $0.866$ & $7.9788\times10^{-5}$ & $-1.4497$ \\
		$1$ & $28$ & $0.11115$ & $0.862$ & $4.5460\times10^{-5}$ & $-1.5117$ \\
		$1$ & $28$ & $0.11563$ & $0.858$ & $0$ & $-1.5811$ \\
		$1$ & $28$ & $0.12173$ & $0.853$ & $-6.8851\times10^{-5}$ & $-1.6806$ \\
		$1$ & $28$ & $0.12856$ & $0.848$ & $-1.6494\times10^{-4}$ & $-1.7993$ \\
		\hline
		
		$1$ & $58$ & $0.02103$ & $0.973$ & $2.4287\times10^{-5}$ & $-0.7394$ \\
		$1$ & $58$ & $0.02818$ & $0.960$ & $4.8074\times10^{-5}$ & $-0.78243$ \\
		$1$ & $58$ & $0.03364$ & $0.949$ & $7.0986\times10^{-5}$ & $-0.8232$ \\
		$1$ & $58$ & $0.04250$ & $0.930$ & $1.1111\times10^{-4}$ & $-0.9049$ \\
		$1$ & $58$ & $0.04800$ & $0.918$ & $1.3333\times10^{-4}$ & $-0.9657$ \\
		$1$ & $58$ & $0.06153$ & $0.890$ & $1.5800\times10^{-4}$ & $-1.1486$ \\
		$1$ & $58$ & $0.06860$ & $0.877$ & $1.4505\times10^{-4}$ & $-1.2634$ \\
		$1$ & $58$ & $0.07529$ & $0.866$ & $1.1237\times10^{-4}$ & $-1.3846$ \\
		$1$ & $58$ & $0.07933$ & $0.860$ & $8.2172\times10^{-5}$ & $-1.4638$ \\
		$1$ & $58$ & $0.08224$ & $0.856$ & $5.5441\times10^{-5}$ & $-1.5235$ \\
		$1$ & $58$ & $0.08698$ & $0.850$ & $0$ & $-1.6262$ \\
		$1$ & $58$ & $0.09048$ & $0.846$ & $-4.4479\times10^{-5}$ & $-1.706$ \\
		$1$ & $58$ & $0.09751$ & $0.839$ & $-1.5881\times10^{-4}$ & $-1.877$ \\
	\end{longtable*}
	
	\clearpage
	
	
	\begin{longtable*}{@{}cccccc@{}}
		\caption{Radial perturbation data for the second excited-state solutions ($n=2$).}
		\label{tab:radial-n2}\\
		\hline\hline
		$n$ & $\eta$ & $\phi(0)$ & $\omega$ & $\chi^2$ & $\zeta$ \\
		\hline\hline
		\endfirsthead
		\multicolumn{6}{c}{\tablename~\thetable\ (continued)}\\
		\hline\hline
		$n$ & $\eta$ & $\phi(0)$ & $\omega$ & $\chi^2$ & $\zeta$ \\
		\hline\hline
		\endhead
		\hline
		\multicolumn{6}{r}{Continued on next page}\\
		\endfoot
		\hline\hline
		\endlastfoot
		
		$2$ & $0$ & $0.04103$ & $0.966$ & $1.5011\times10^{-5}$ & $-0.77393$ \\
		$2$ & $0$ & $0.06572$ & $0.948$ & $3.0223\times10^{-5}$ & $-0.85407$ \\
		$2$ & $0$ & $0.08686$ & $0.934$ & $4.1326\times10^{-5}$ & $-0.93405$ \\
		$2$ & $0$ & $0.09991$ & $0.926$ & $4.5818\times10^{-5}$ & $-0.98936$ \\
		$2$ & $0$ & $0.10849$ & $0.921$ & $4.7393\times10^{-5}$ & $-1.0285$ \\
		$2$ & $0$ & $0.12302$ & $0.913$ & $4.7010\times10^{-5}$ & $-1.1001$ \\
		$2$ & $0$ & $0.14927$ & $0.900$ & $3.4382\times10^{-5}$ & $-1.2486$ \\
		$2$ & $0$ & $0.16049$ & $0.895$ & $2.3436\times10^{-5}$ & $-1.3206$ \\
		$2$ & $0$ & $0.17760$ & $0.888$ & $0$ & $-1.4414$ \\
		$2$ & $0$ & $0.18555$ & $0.885$ & $-1.5156\times10^{-5}$ & $-1.5025$ \\
		$2$ & $0$ & $0.22280$ & $0.873$ & $-1.1546\times10^{-4}$ & $-1.8368$ \\
		\hline
		
		$2$ & $7.9577$ & $0.03350$ & $0.970$ & $1.2189\times10^{-5}$ & $-0.75924$ \\
		$2$ & $7.9577$ & $0.04481$ & $0.960$ & $2.0313\times10^{-5}$ & $-0.79895$ \\
		$2$ & $7.9577$ & $0.05868$ & $0.948$ & $3.1265\times10^{-5}$ & $-0.85449$ \\
		$2$ & $7.9577$ & $0.08049$ & $0.930$ & $4.6814\times10^{-5}$ & $-0.95912$ \\
		$2$ & $7.9577$ & $0.09341$ & $0.920$ & $5.2596\times10^{-5}$ & $-1.0323$ \\
		$2$ & $7.9577$ & $0.10017$ & $0.915$ & $5.3977\times10^{-5}$ & $-1.0743$ \\
		$2$ & $7.9577$ & $0.11006$ & $0.908$ & $5.3433\times10^{-5}$ & $-1.1405$ \\
		$2$ & $7.9577$ & $0.11749$ & $0.903$ & $5.0699\times10^{-5}$ & $-1.1942$ \\
		$2$ & $7.9577$ & $0.13182$ & $0.894$ & $3.8802\times10^{-5}$ & $-1.3084$ \\
		$2$ & $7.9577$ & $0.14786$ & $0.885$ & $1.3488\times10^{-5}$ & $-1.4535$ \\
		$2$ & $7.9577$ & $0.15391$ & $0.8819$ & $0$ & $-1.5134$ \\
		$2$ & $7.9577$ & $0.15988$ & $0.879$ & $-1.5059\times10^{-5}$ & $-1.5756$ \\
		$2$ & $7.9577$ & $0.18103$ & $0.870$ & $-8.8154\times10^{-5}$ & $-1.8207$ \\
		\hline
		
		$2$ & $28$ & $0.02798$ & $0.971$ & $1.1644\times10^{-5}$ & $-0.75633$ \\
		$2$ & $28$ & $0.03792$ & $0.959$ & $2.1624\times10^{-5}$ & $-0.80392$ \\
		$2$ & $28$ & $0.04901$ & $0.945$ & $3.5038\times10^{-5}$ & $-0.86923$ \\
		$2$ & $28$ & $0.05603$ & $0.936$ & $4.3735\times10^{-5}$ & $-0.91798$ \\
		$2$ & $28$ & $0.06469$ & $0.925$ & $5.3250\times10^{-5}$ & $-0.98647$ \\
		$2$ & $28$ & $0.07032$ & $0.918$ & $5.7987\times10^{-5}$ & $-1.0363$ \\
		$2$ & $28$ & $0.07695$ & $0.910$ & $6.1348\times10^{-5}$ & $-1.1005$ \\
		$2$ & $28$ & $0.08125$ & $0.905$ & $6.1912\times10^{-5}$ & $-1.1454$ \\
		$2$ & $28$ & $0.08570$ & $0.900$ & $6.0921\times10^{-5}$ & $-1.1947$ \\
		$2$ & $28$ & $0.09619$ & $0.889$ & $5.1154\times10^{-5}$ & $-1.3225$ \\
		$2$ & $28$ & $0.10812$ & $0.878$ & $2.5178\times10^{-5}$ & $-1.4883$ \\
		$2$ & $28$ & $0.11555$ & $0.872$ & $0$ & $-1.6031$ \\
		$2$ & $28$ & $0.11960$ & $0.869$ & $-1.7120\times10^{-5}$ & $-1.6697$ \\
		$2$ & $28$ & $0.133822$ & $0.860$ & $-9.6088\times10^{-5}$ & $-1.9252$ \\
		\hline
		
		$2$ & $58$ & $0.02533$ & $0.968$ & $1.3853\times10^{-5}$ & $-0.76795$ \\
		$2$ & $58$ & $0.03277$ & $0.955$ & $2.4953\times10^{-5}$ & $-0.82057$ \\
		$2$ & $58$ & $0.04124$ & $0.939$ & $4.0009\times10^{-5}$ & $-0.89642$ \\
		$2$ & $58$ & $0.05108$ & $0.920$ & $5.6035\times10^{-5}$ & $-1.0073$ \\
		$2$ & $58$ & $0.05642$ & $0.910$ & $6.1695\times10^{-5}$ & $-1.0781$ \\
		$2$ & $58$ & $0.06261$ & $0.899$ & $6.3781\times10^{-5}$ & $-1.1695$ \\
		$2$ & $58$ & $0.06497$ & $0.895$ & $6.3025\times10^{-5}$ & $-1.2071$ \\
		$2$ & $58$ & $0.06804$ & $0.890$ & $6.0592\times10^{-5}$ & $-1.2583$ \\
		$2$ & $58$ & $0.07757$ & $0.876$ & $4.1405\times10^{-5}$ & $-1.4333$ \\
		$2$ & $58$ & $0.08307$ & $0.869$ & $2.1497\times10^{-5}$ & $-1.5458$ \\
		$2$ & $58$ & $0.08744$ & $0.864$ & $0$ & $-1.6415$ \\
		$2$ & $58$ & $0.09233$ & $0.859$ & $-2.8025\times10^{-5}$ & $-1.7548$ \\
		$2$ & $58$ & $0.09674$ & $0.855$ & $-5.9071\times10^{-5}$ & $-1.863$ \\
	\end{longtable*}
	
	\clearpage
	
	
	\begin{longtable*}{@{}cccccc@{}}
		\caption{Radial perturbation data for the third excited-state solutions ($n=3$).}
		\label{tab:radial-n3}\\
		\hline\hline
		$n$ & $\eta$ & $\phi(0)$ & $\omega$ & $\chi^2$ & $\zeta$ \\
		\hline\hline
		\endfirsthead
		\multicolumn{6}{c}{\tablename~\thetable\ (continued)}\\
		\hline\hline
		$n$ & $\eta$ & $\phi(0)$ & $\omega$ & $\chi^2$ & $\zeta$ \\
		\hline\hline
		\endhead
		\hline
		\multicolumn{6}{r}{Continued on next page}\\
		\endfoot
		\hline\hline
		\endlastfoot
		
		$3$ & $0$ & $0.05016$ & $0.960$ & $1.0709\times10^{-5}$ & $-0.81158$ \\
		$3$ & $0$ & $0.06902$ & $0.947$ & $1.6680\times10^{-5}$ & $-0.88059$ \\
		$3$ & $0$ & $0.08801$ & $0.935$ & $2.1639\times10^{-5}$ & $-0.95986$ \\
		$3$ & $0$ & $0.09820$ & $0.929$ & $2.3473\times10^{-5}$ & $-1.0069$ \\
		$3$ & $0$ & $0.11456$ & $0.920$ & $2.4737\times10^{-5}$ & $-1.0899$ \\
		$3$ & $0$ & $0.12431$ & $0.915$ & $2.4306\times10^{-5}$ & $-1.144$ \\
		$3$ & $0$ & $0.14336$ & $0.906$ & $2.0441\times10^{-5}$ & $-1.2609$ \\
		$3$ & $0$ & $0.15496$ & $0.901$ & $1.5872\times10^{-5}$ & $-1.3401$ \\
		$3$ & $0$ & $0.17922$ & $0.8916$ & $0$ & $-1.5278$ \\
		$3$ & $0$ & $0.21759$ & $0.880$ & $-4.4866\times10^{-5}$ & $-1.8968$ \\
		$3$ & $0$ & $0.25426$ & $0.872$ & $-1.1730\times10^{-4}$ & $-2.352$ \\
		\hline
		
		$3$ & $7.9577$ & $0.04607$ & $0.960$ & $1.0885\times10^{-5}$ & $-0.812$ \\
		$3$ & $7.9577$ & $0.05936$ & $0.949$ & $1.6158\times10^{-5}$ & $-0.869$ \\
		$3$ & $7.9577$ & $0.07321$ & $0.938$ & $2.1297\times10^{-5}$ & $-0.93714$ \\
		$3$ & $7.9577$ & $0.09064$ & $0.925$ & $2.5901\times10^{-5}$ & $-1.037$ \\
		$3$ & $7.9577$ & $0.09775$ & $0.920$ & $2.6821\times10^{-5}$ & $-1.0828$ \\
		$3$ & $7.9577$ & $0.10516$ & $0.915$ & $2.7011\times10^{-5}$ & $-1.1336$ \\
		$3$ & $7.9577$ & $0.11290$ & $0.910$ & $2.6243\times10^{-5}$ & $-1.1907$ \\
		$3$ & $7.9577$ & $0.12973$ & $0.900$ & $2.0587\times10^{-5}$ & $-1.3287$ \\
		$3$ & $7.9577$ & $0.14095$ & $0.894$ & $1.3315\times10^{-5}$ & $-1.4326$ \\
		$3$ & $7.9577$ & $0.15469$ & $0.8874$ & $0$ & $-1.5735$ \\
		$3$ & $7.9577$ & $0.16743$ & $0.882$ & $-1.6928\times10^{-5}$ & $-1.7189$ \\
		$3$ & $7.9577$ & $0.17810$ & $0.878$ & $-3.5035\times10^{-5}$ & $-1.8525$ \\
		\hline
		
		$3$ & $28$ & $0.03671$ & $0.962$ & $1.0152\times10^{-5}$ & $-0.80314$ \\
		$3$ & $28$ & $0.04520$ & $0.952$ & $1.5061\times10^{-5}$ & $-0.85212$ \\
		$3$ & $28$ & $0.05526$ & $0.940$ & $2.1159\times10^{-5}$ & $-0.92075$ \\
		$3$ & $28$ & $0.06806$ & $0.925$ & $2.7691\times10^{-5}$ & $-1.0262$ \\
		$3$ & $28$ & $0.07700$ & $0.915$ & $3.0176\times10^{-5}$ & $-1.1129$ \\
		$3$ & $28$ & $0.08168$ & $0.910$ & $3.0449\times10^{-5}$ & $-1.1628$ \\
		$3$ & $28$ & $0.08654$ & $0.905$ & $2.9821\times10^{-5}$ & $-1.2179$ \\
		$3$ & $28$ & $0.09375$ & $0.898$ & $2.6930\times10^{-5}$ & $-1.3062$ \\
		$3$ & $28$ & $0.10396$ & $0.889$ & $1.8237\times10^{-5}$ & $-1.4449$ \\
		$3$ & $28$ & $0.11576$ & $0.880$ & $0$ & $-1.6261$ \\
		$3$ & $28$ & $0.12337$ & $0.875$ & $-1.5735\times10^{-5}$ & $-1.7552$ \\
		$3$ & $28$ & $0.13212$ & $0.870$ & $-3.9593\times10^{-5}$ & $-1.9164$ \\
		\hline
		
		$3$ & $58$ & $0.03061$ & $0.961$ & $1.0626\times10^{-5}$ & $-0.80737$ \\
		$3$ & $58$ & $0.03706$ & $0.950$ & $1.6065\times10^{-5}$ & $-0.86042$ \\
		$3$ & $58$ & $0.04551$ & $0.935$ & $2.3641\times10^{-5}$ & $-0.94535$ \\
		$3$ & $58$ & $0.05115$ & $0.925$ & $2.8064\times10^{-5}$ & $-1.0122$ \\
		$3$ & $58$ & $0.05637$ & $0.916$ & $3.1014\times10^{-5}$ & $-1.0814$ \\
		$3$ & $58$ & $0.06059$ & $0.909$ & $3.2232\times10^{-5}$ & $-1.1426$ \\
		$3$ & $58$ & $0.06309$ & $0.905$ & $3.2347\times10^{-5}$ & $-1.1812$ \\
		$3$ & $58$ & $0.06633$ & $0.900$ & $3.1736\times10^{-5}$ & $-1.2336$ \\
		$3$ & $58$ & $0.07257$ & $0.891$ & $2.7824\times10^{-5}$ & $-1.3427$ \\
		$3$ & $58$ & $0.07634$ & $0.886$ & $2.3536\times10^{-5}$ & $-1.4139$ \\
		$3$ & $58$ & $0.08212$ & $0.879$ & $1.3895\times10^{-5}$ & $-1.5306$ \\
		$3$ & $58$ & $0.08770$ & $0.873$ & $0$ & $-1.6523$ \\
		$3$ & $58$ & $0.09298$ & $0.868$ & $-1.5139\times10^{-5}$ & $-1.7758$ \\
		$3$ & $58$ & $0.09778$ & $0.864$ & $-3.2872\times10^{-5}$ & $-1.8951$ \\
	\end{longtable*}
	
	\endgroup
	\twocolumngrid
	\clearpage

	\bibliographystyle{apsrev4-2}

\end{document}